\documentclass[preprint2]{aastex}

\usepackage{graphicx}
\DeclareGraphicsExtensions{.ps}
\shorttitle{\indent \def Are IRIS bombs Ellerman bombs?} \shortauthors{Tian et al.}

\begin{document}

\title{Are IRIS bombs connected to Ellerman bombs?}

\author{Hui Tian\altaffilmark{1,2}, Zhi Xu\altaffilmark{3}, Jiansen He\altaffilmark{1}, Chad Madsen\altaffilmark{4,5}}
\altaffiltext{1}{School of Earth and Space Sciences, Peking University, Beijing 100871, China; huitian@pku.edu.cn}
\altaffiltext{2}{State Key Laboratory of Space Weather, Chinese Academy of Sciences, Beijing 100190, China}
\altaffiltext{3}{Yunnan Astronomical Observatory, Chinese Academy of Sciences, Kunming 650011, China}
\altaffiltext{4}{Harvard-Smithsonian Center for Astrophysics, 60 Garden Street, Cambridge, MA 02138, USA}
\altaffiltext{5}{Center for Space Physics, Boston University, 725 Commonwealth Ave., Boston, MA 02215, USA}

\begin{abstract}
Recent observations by the Interface Region Imaging Spectrograph (IRIS) have revealed pockets of hot gas ($\sim$2--8$\times$10$^{4}$ K) potentially resulting from magnetic reconnection in the partially ionized lower solar atmosphere (IRIS bombs; IBs). Using joint observations between IRIS and the Chinese New Vacuum Solar Telescope, we have identified ten IBs. We find that three are unambiguously and three others are possibly connected to Ellerman bombs (EBs), which show intense brightening of the extended H$_{\alpha}$ wings without leaving an obvious signature in the H$_{\alpha}$ core. These bombs generally reveal the following distinct properties: (1) The O~{\sc{iv}}~1401.156\AA{} and 1399.774\AA{} lines are absent or very weak; (2) The Mn~{\sc{i}}~2795.640\AA{} line manifests as an absorption feature superimposed on the greatly enhanced Mg~{\sc{ii}}~k line wing; (3) The Mg~{\sc{ii}}~k and h lines show intense brightening in the wings and no dramatic enhancement in the cores; (4) Chromospheric absorption lines such as Ni~{\sc{ii}}~1393.330\AA{} and 1335.203\AA{} are very strong; (5) The 1700\AA{} images obtained with the Atmospheric Imaging Assembly on board the Solar Dynamics Observatory reveal intense and compact brightenings. These properties support the formation of these bombs in the photosphere, demonstrating that EBs can be heated much more efficiently than previously thought. We also demonstrate that the Mg~{\sc{ii}}~k and h lines can be used to investigate EBs similarly to H$_{\alpha}$, which opens a promising new window for EB studies. The remaining four IBs obviously have no connection to EBs and they do not have the properties mentioned above, suggesting a higher formation layer possibly in the chromosphere. 
\end{abstract}

\section{Introduction}

Recent high-resolution observations by the Interface Region Imaging Spectrograph \citep[e.g.,][]{DePontieu2014a} have provided fascinating new insights into the energetics of the lower solar atmosphere \citep[e.g.,][]{Peter2014,Tian2014a,DePontieu2014b,Hansteen2014,Testa2014}. One new finding is the discovery of absorption lines (singly ionized or neutral) superimposed on the greatly broadened transition region lines \citep{Peter2014}. Such line profiles are typically found in small-scale compact bright regions from slit-jaw images (SJI) taken with the 1400\AA{} and 1330\AA{} filters, and they were suggested to indicate local heating of the photosphere or lower chromosphere to $\sim$8$\times$10$^{4}$ K under the assumption of collisional ionization equilibrium. However, \cite{Judge2015} performed an independent analysis of the same dataset and suggested that these events arise from plasma originally at pressure between $\leqslant$80 and 800 dyne cm$^{-2}$, which places the origin of these events in the low-middle chromosphere or above. In this paper we call these evens IRIS bombs (IBs). 

\cite{Peter2014} mentioned that these IBs may be connected to Ellerman bombs (EBs), which were discovered by \cite{Ellerman1917} and characterized as intense short-lived brightening of the extended wings of the H$_{\alpha}$ line at 6563\AA{}. These events were first called "solar hydrogen bombs" and renamed as EBs by \cite{McMath1960}. \cite{Pariat2007} and \cite{SocasNavarro2006} found that EBs can also be observed in the Ca~{\sc{ii}}~8542\AA{} line. Although some studies put the formation layer of EBs in the low chromosphere \citep[e.g.,][]{Schmieder2004,Yang2013}, recent high-resolution observations suggested that EBs are a purely photospheric phenomenon and that they often reveal upright flame morphology in limbward viewing \citep[e.g.,][]{Watanabe2011,Nelson2015}. Recently, \cite{Rutten2013} and \cite{Vissers2015} pointed out that many previously identified EBs are likely "pseudo-EBs", which are the much more ubiquitous facular or network bright points and their brightenings in H$_{\alpha}$ wings are usually not as significant as those of EBs \citep{Watanabe2011}. These "pseudo-EBs" are indicators of deeper-than-normal radiation escape rather than magnetic reconnection. It is suggested that at least some EBs mark anti-parallel reconnection in the photosphere during the emergence of active regions \citep[e.g.,][]{Vissers2013,Reid2016}. Modeling of the H$_{\alpha}$ wing enhancement generally indicates a temperature increase by a few hundred to $\sim$3000 Kelvin in the upper photosphere or lower chromosphere \citep[e.g.,][]{Fang2006,Isobe2007,BelloGonzalez2013,Berlicki2014,Hong2014}. For details about the morphology and properties of EBs, we refer to the reviews of \cite{Georgoulis2002} and \cite{Rutten2013}.

Since EBs are defined from H$_{\alpha}$ observations, simultaneous observations of IRIS and an H$_{\alpha}$ instrument are required to investigate the relationship between IBs and EBs. Using data taken by the Swedish 1-m Solar Telescope \citep[SST,][]{Scharmer2003}, \cite{Vissers2015} studied five EBs and concluded that strong EB activity can indeed produce IB-type spectra. Based on the joint observations between IRIS and the New Solar Telescope\citep[NST,][]{Cao2010} on 2014 July 30, \cite{Kim2015} identified the connection between an obvious IB and a weak EB. Apparently, more coordinated observations between IRIS and H$_{\alpha}$ instruments need to be performed to investigate the relationship between IBs and EBs. This type of observations can also provide important constraint to numerical simulations of magnetic reconnection in the partially ionized lower solar atmosphere \citep[e.g.,][]{Murphy2015,Ni2015}. 

Here we use joint observations between IRIS and the Chinese 1-m New Vacuum Solar Telescope \citep[NVST,][] {Liu2014} to examine the possible connection between IBs and EBs. The NVST belongs to a new generation of large and high-technology solar facilities of China and one of the post-focus instruments is the Multi-channel High Resolution Imaging System, including H$_{\alpha}$, G-band, TiO band, Ca~{\sc{ii}}~8542\AA{} and He~{\sc{i}}~10830\AA{} wavelengths \citep{Xu2013}. In the present paper, among the data taken with the NVST, we only report results obtained in the H$_{\alpha}$ channel, of which the central wavelength can be tunable in the range of 6562.8 +/-- 4\AA{} and the full bandpass width is 0.25\AA{}. Our investigation clearly reveals that some IBs are connected to EBs and others are not.

\section{Observations}

\begin{figure*}
\centering {\includegraphics[width=\textwidth]{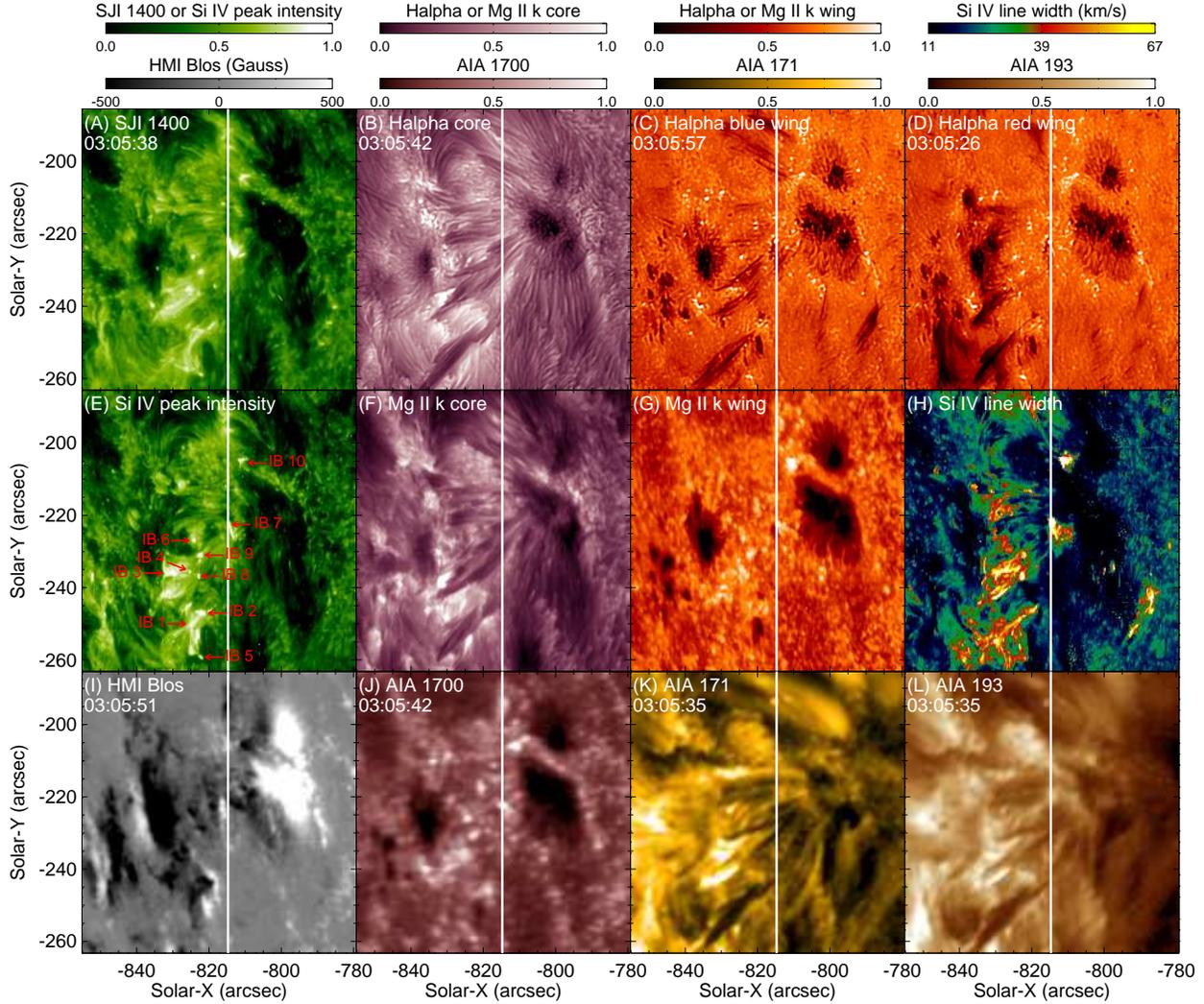}} \caption{ (A)-(D) IRIS/SJI 1400\AA{} image, NVST H$_{\alpha}$ core and wing (--1\AA{} and +1\AA{}) images taken around 03:05:38 UT. The dark filamentary structures in the H$_{\alpha}$ wings, especially in the blue wing, are chromospheric spicules which could affect the detection of EBs. (E)-(H) Images of the Si~{\sc{iv}}~1393.755\AA{}~intensity, Mg~{\sc{ii}}~k core and wing (sum of --1.33\AA{} and +1.33\AA{}), and Si~{\sc{iv}}~1393.755\AA{}~line width. (I)-(L) SDO/HMI line-of-sight magnetogram, SDO/AIA 1700\AA{}, 171\AA{} and 193\AA{} images taken around 03:05:38 UT. The white line in each panel indicates the slit location at the corresponding time. Two movies (m1.mov and m2.mov) showing the IRIS, NVST and SDO observations are available online. Ten IBs are indicated by the red arrows in panel (E). } \label{fig.1}
\end{figure*}

The joint IRIS and NVST observations were performed on 2015 May 2. IRIS performed a very large dense raster (175$^{\prime\prime}$ along the slit, 400 raster steps with a step size of $\sim$0.33$^{\prime\prime}$) of the emerging NOAA active region (AR) 12335 from 02:34 UT to 03:36 UT. The pointing coordinate was (--814$^{\prime\prime}$, --222$^{\prime\prime}$), close to the east limb. The data was summed onboard by 2 both spectrally and spatially, leading to a spatial pixel size of $\sim$0.33$^{\prime\prime}$ and a spectral dispersion of $\sim$0.026 \AA{}/$\sim$0.051 \AA{} per pixel in the far/near ultraviolet wavelength bands. The cadence of the spectral observation was $\sim$9.2 seconds, with an exposure time of 8 seconds. Slit-jaw images (SJI) in the 1400\AA{}~(mainly ultraviolet continuum and Si~{\sc{iv}}), 1330\AA{}~(mainly ultraviolet continuum and C~{\sc{ii}}) and 2796\AA{}~(mainly Mg~{\sc{ii}} k) filters were taken with a cadence of $\sim$36.7 seconds for each filter. Dark current subtraction, flat field, geometrical and orbital variation corrections have been applied in the level 2 data used here. The fiducial lines are used to achieve a coalignment between different SJI filters and different spectral windows. The SJI images are internally coaligned by removing the solar rotation effect.

The NVST observation lasted from 01:11 UT to 03:59 UT. We took images of the H$_{\alpha}$ line core, blue wing at --1\AA{} and red wing at +1\AA{} alternately, with a cadence of $\sim$52 seconds for each filter. These images have a spatial pixel size of $\sim$0.167$^{\prime\prime}$ and a field of view of $\sim$153$^{\prime\prime}$$\times$153$^{\prime\prime}$. The data reduction of the dark current and flat field modification was followed by the image reconstruction precess based on the speckle masking method \citep{Weigelt1977,Liu1998}. We rotate the NVST images by 54.28$^{\circ}$ so that the vertical dimension of the images is oriented in the north-south direction, same as the IRIS images. The coalignment between IRIS images and NVST images in different filters is achieved by doing the following: We first build a Mg~{\sc{ii}}~k core image and a wing (sum of --1.33\AA{} and +1.33\AA{}) image from the IRIS spectral data taken at different times. Bright features visible in the Mg~{\sc{ii}}~k core/wing image are then compared with the associated bright dynamic features in H$_{\alpha}$ core/wing image sequences.

To investigate the response of IRIS bombs at different temperatures and study the magnetic field structures associated with these bombs, we have also analyzed the data taken by the Atmospheric Imaging Assembly \citep[AIA,][]{Lemen2012} and the Helioseismic and Magnetic Imager \citep[HMI,][]{Scherrer2012} onboard the Solar Dynamics Observatory \citep[SDO,][]{Pesnell2012}. The AIA images were taken with a cadence of 12 seconds in the 171\AA{} and 193\AA{} passbands and 24 seconds in the 1700\AA{} passband. The cadence of the line-of-sight magnetograms taken by HMI is 45 seconds. The pixel sizes of the AIA and HMI images are $\sim$0.613$^{\prime\prime}$ and $\sim$0.504$^{\prime\prime}$, respectively. We coalign the AIA 1700\AA{}~(mainly ultraviolet continuum formed around the temperature minimum) and IRIS 1400\AA{}~images by checking locations of the commonly observed sunspots and some bright features. The AIA 171\AA{} and 193\AA{} images should then be automatically aligned with the IRIS images since AIA images in different passbands are automatically coaligned after applying the standard SolarSoft (SSW) routine aia\_prep.pro. The coalignment between HMI magnetograms and IRIS images is achieved by matching the bright network lanes in 1400\AA{}~images and the flux concentrations in magnetograms immediately outside the AR. 

Figure~\ref{fig.1} presents the IRIS/SJI 1400\AA{} image, NVST H$_{\alpha}$ core and wing images, HMI magnetogram, AIA 1700\AA{}, 171\AA{} and 193\AA{} images taken around 03:05:38 UT. Also shown are the Mg~{\sc{ii}}~k core and wing images, as well as the intensity and line width images obtained by applying a single Gaussian fit to the Si~{\sc{iv}}~1393.755\AA{}~line profiles. The region shown here represents only part of the full field of view of IRIS. Time sequences of these images are presented in two online movies. To demonstrate the good coalignment of images taken by different instruments, contours of the AIA 1700\AA{} images are also overplotted in one movie.  

\begin{figure*}
\centering {\includegraphics[width=\textwidth]{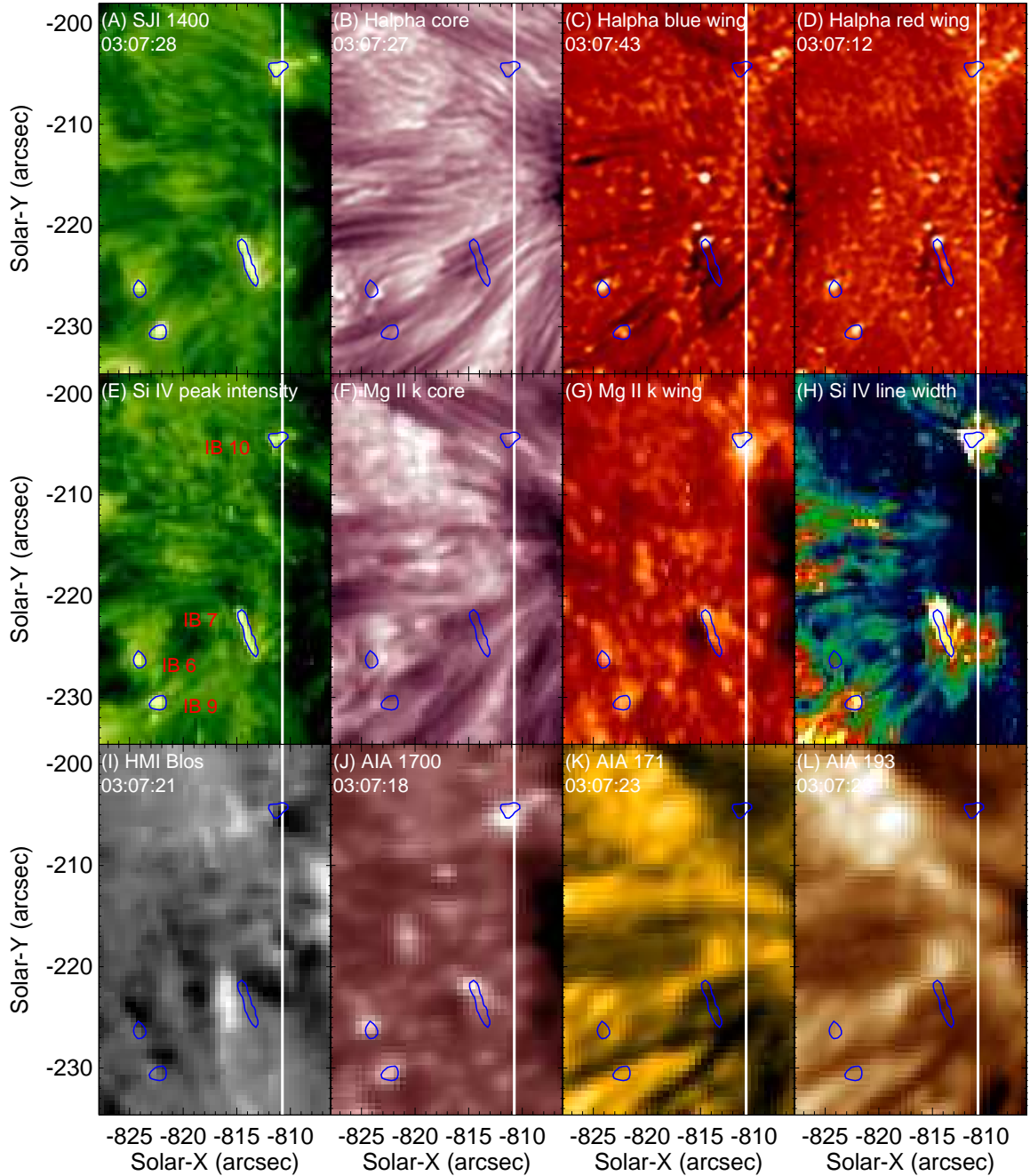}} \caption{ Similar to Figure~\ref{fig.1} but only a small region enclosing four IBs is shown. Locations of the IBs are marked by overplotting contours of the Si~{\sc{iv}}~1393.755\AA{}~peak intensity. The white line in each panel indicates the slit location at the time 03:07:28 UT. A movie (m3.mov) showing the IRIS, NVST and SDO observations is available online. } \label{fig.2}
\end{figure*}

To take a closer look at a few IBs we also show only a small region around the coordinate of (--817$^{\prime\prime}$, --216$^{\prime\prime}$) in Figure~\ref{fig.2} and the associated online movie. Four IBs have been identified in this region (see below). The online movie reveals some dark filamentary structures in the H$_{\alpha}$ wings, especially in the blue wing. As we will show below, these long structures are presumably chromospheric spicules and they could affect the detection of EBs.

\section{Results and Discussion}
\subsection{Identification of IRIS bombs}
We identify IBs based mainly on the far ultraviolet spectra. We first select all substantial and compact brightenings in the Si~{\sc{iv}}~1393.755\AA{}~intensity image. Line profiles in these brightenings are then inspected. IBs are defined as those brightenings with chromospheric absorption lines superimposed on greatly broadened and enhanced non-Gaussian profiles of transition region lines (e.g., Si~{\sc{iv}} and C~{\sc{ii}}). The absorption feature needs to be obvious for at least the Ni~{\sc{ii}}~1393.330\AA{} and 1335.203\AA{} lines. Based on this criterion, we have identified ten IBs sampled by the slit and their locations are indicated in Figure~\ref{fig.1}(E).

\begin{figure*}
\centering {\includegraphics[width=\textwidth]{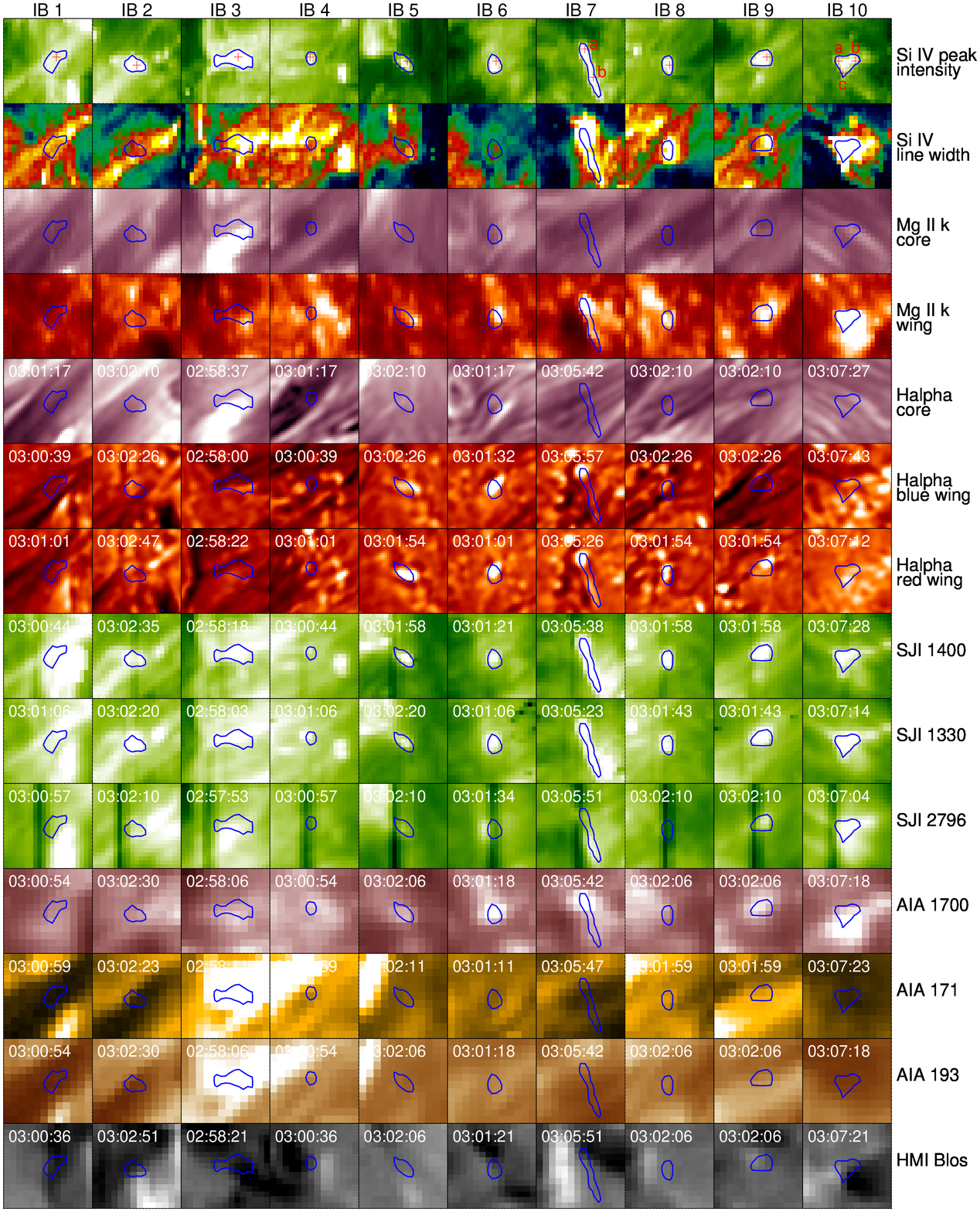}} \caption{ IRIS, NVST and SDO images showing a 7$^{\prime\prime}$$\times$7$^{\prime\prime}$region enclosing each identified IB. Locations of the IBs are marked by overplotting contours of the Si~{\sc{iv}}~1393.755\AA{}~peak intensity. The observing times of the IRIS/SJI, NVST and SDO images are also marked in corresponding panels. The red crosses shown in the Si~{\sc{iv}}~intensity images indicate the locations where the line profiles presented in Figures~\ref{fig.4}--\ref{fig.9} are obtained. } \label{fig.3}
\end{figure*}

Figure~\ref{fig.3} presents images of the Si~{\sc{iv}}~1393.755\AA{}~intensity and line width, Mg~{\sc{ii}}~k core and wing, SJI 1400\AA{}, 1330\AA{} and 2796\AA{}, H$_{\alpha}$ core and wings, line of sight component of the photospheric magnetic field, AIA 1700\AA{}, 171\AA{} and 193\AA{} in a small region enclosing each identified IB. We can see that all IBs are associated with enhanced line width of the Si~{\sc{iv}}~line. IBs are also clearly observed in both the SJI 1400\AA{} and 1330\AA{} images, which can be understood since the Si~{\sc{iv}} and C~{\sc{ii}} lines are usually greatly enhanced in IBs. Signatures of IBs are less obvious in the SJI 2796\AA{} images, suggesting that the Mg~{\sc{ii}} k line does not always have a relevant response.

The four IBs identified by \cite{Peter2014} are all associated with mixed magnetic field polarities and at least one of them show clear signature of flux cancelation during the observation. \cite{Vissers2015} also found association of a few IBs with strong opposite-polarity fluxes. The HMI data presented in Figure~\ref{fig.3} reveals a similar pattern: most IBs appear to be sitting at the magnetic field polarity inversion lines. This likely suggests energization of these IBs through interaction between strong fluxes with opposite polarities, likely anti-parallel magnetic reconnection. The association with strong opposite-polarity fluxes is not obvious for IBs 5, 8 and 9, which might be consistent with the third of the three scenarios proposed by \cite{Georgoulis2002}. This scenario involves interaction between the emerging vertical fields and preexisting horizontal fields. In this case the line of sight components of the interacting magnetic fluxes are not necessarily opposite in polarity. However, it might also be related to the line of sight effect as the observed region is close to the limb.

We also try to identify possible coronal signatures of IBs from both the IRIS spectra and AIA images. We find no obvious emission of the IBs in both the Fe~{\sc{xii}}~1349.38\AA{} and Fe~{\sc{xxi}}~1354.08\AA{} lines (not shown here). We realize that both lines are forbidden lines, so that the absence of these lines might be caused by the large density in the source regions of IBs. However, AIA observations in  the 171\AA{} (dominated by emission from Fe~{\sc{ix}}/Fe~{\sc{x}}) and 193\AA{} (dominated by emission from Fe~{\sc{xii}}) passbands also reveal no obvious brightening at locations of most IBs. A possible exception is IB 3, where a very intense loop-like brightening can be identified from both the 171\AA{} and 193\AA{} images. We thus conclude that IBs are generally not heated to coronal temperatures.

\begin{figure*}
\centering
\begin{minipage}[t]{0.8\textwidth}
{\includegraphics[width=\textwidth]{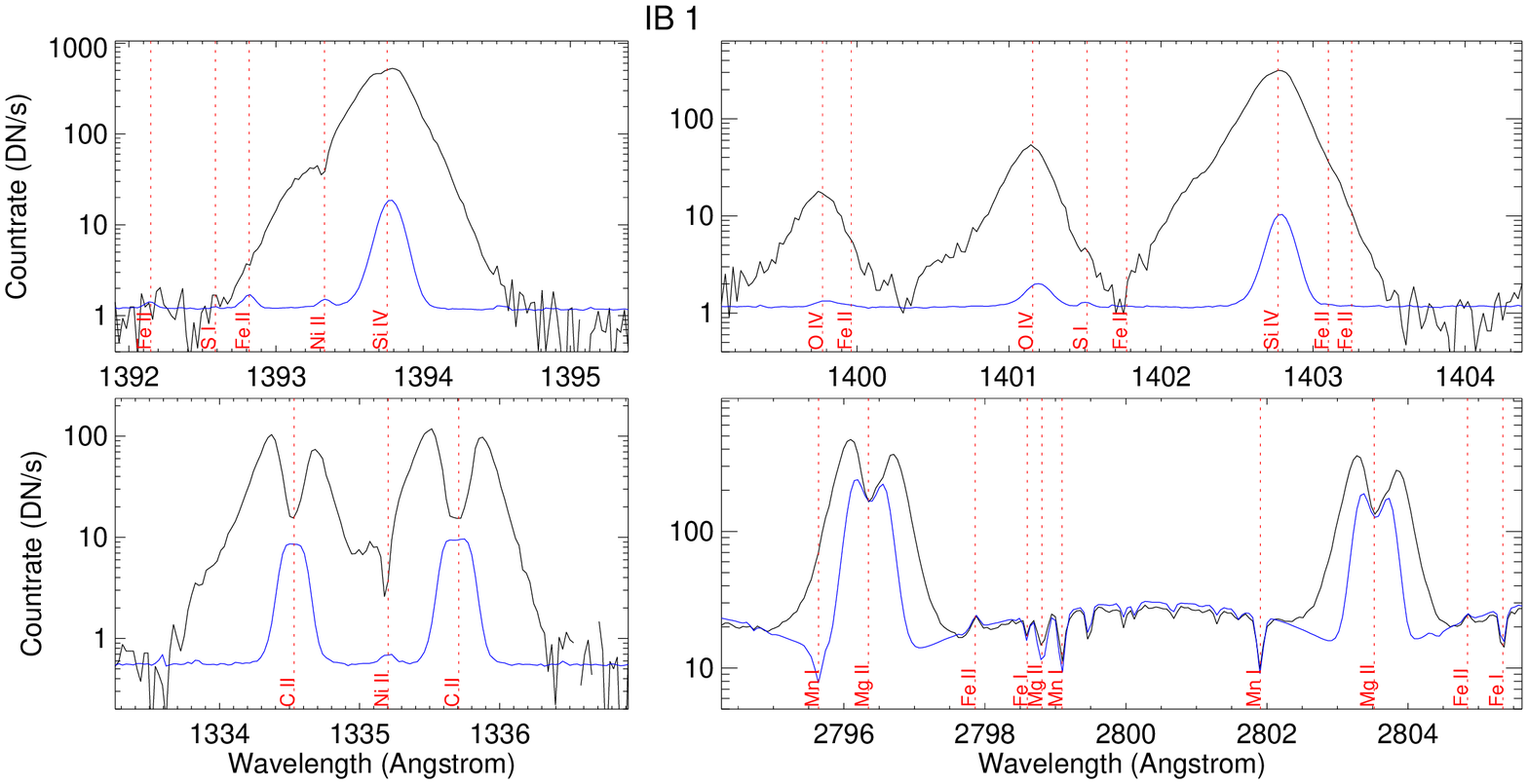}}
\end{minipage}
\begin{minipage}[t]{0.8\textwidth}
{\includegraphics[width=\textwidth]{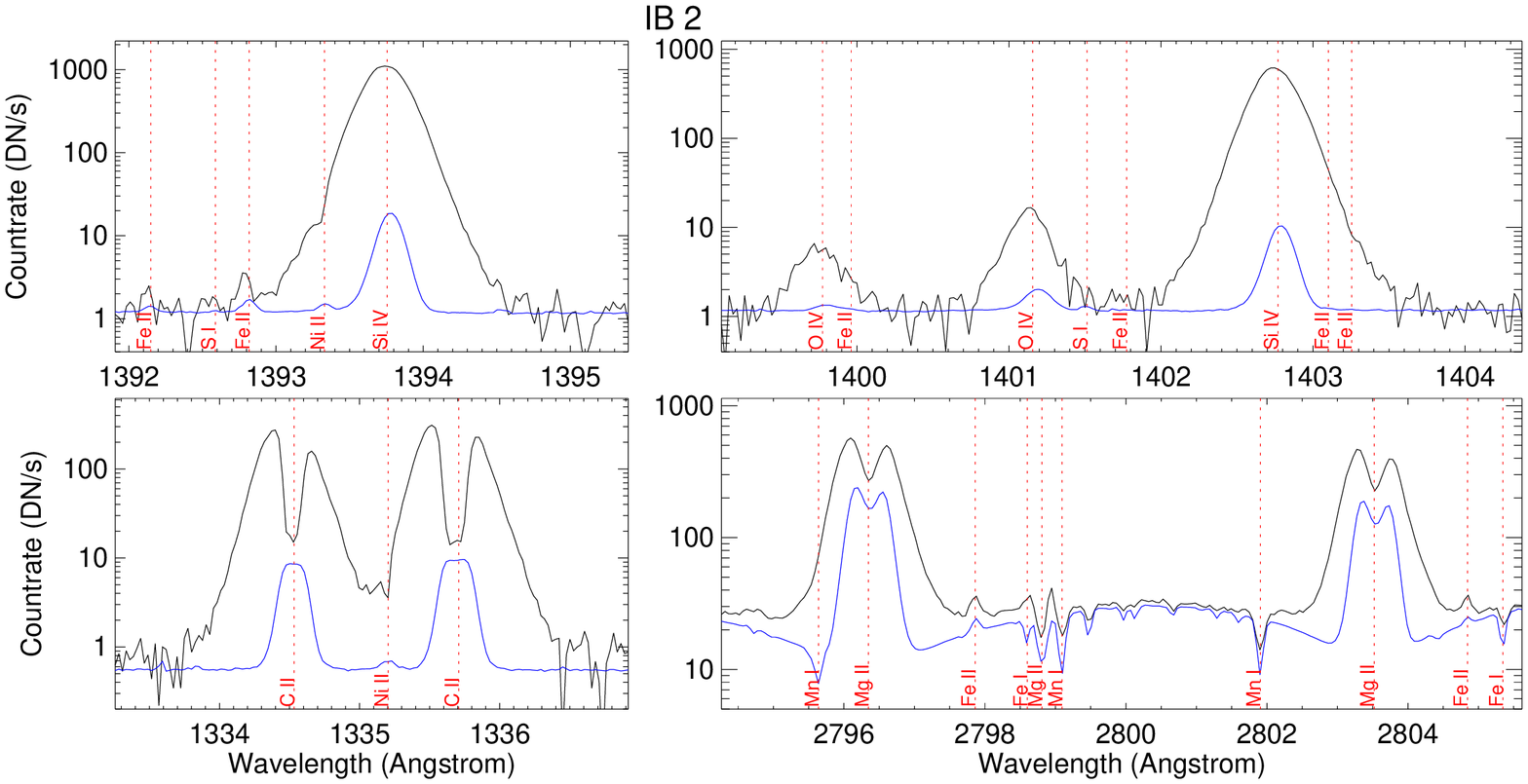}}
\end{minipage}
\begin{minipage}[b]{\textwidth}
\caption{ Typical IRIS line profiles (black lines) of IBs 1 and 2 in four spectral windows. The blue line profiles represent the reference spectra obtained in a quiet plage region. Rest wavelengths of some lines are indicated by the vertical dotted lines. } \label{fig.4}
\end{minipage}
\end{figure*}

\begin{figure*}
\centering
\begin{minipage}[t]{0.8\textwidth}
{\includegraphics[width=\textwidth]{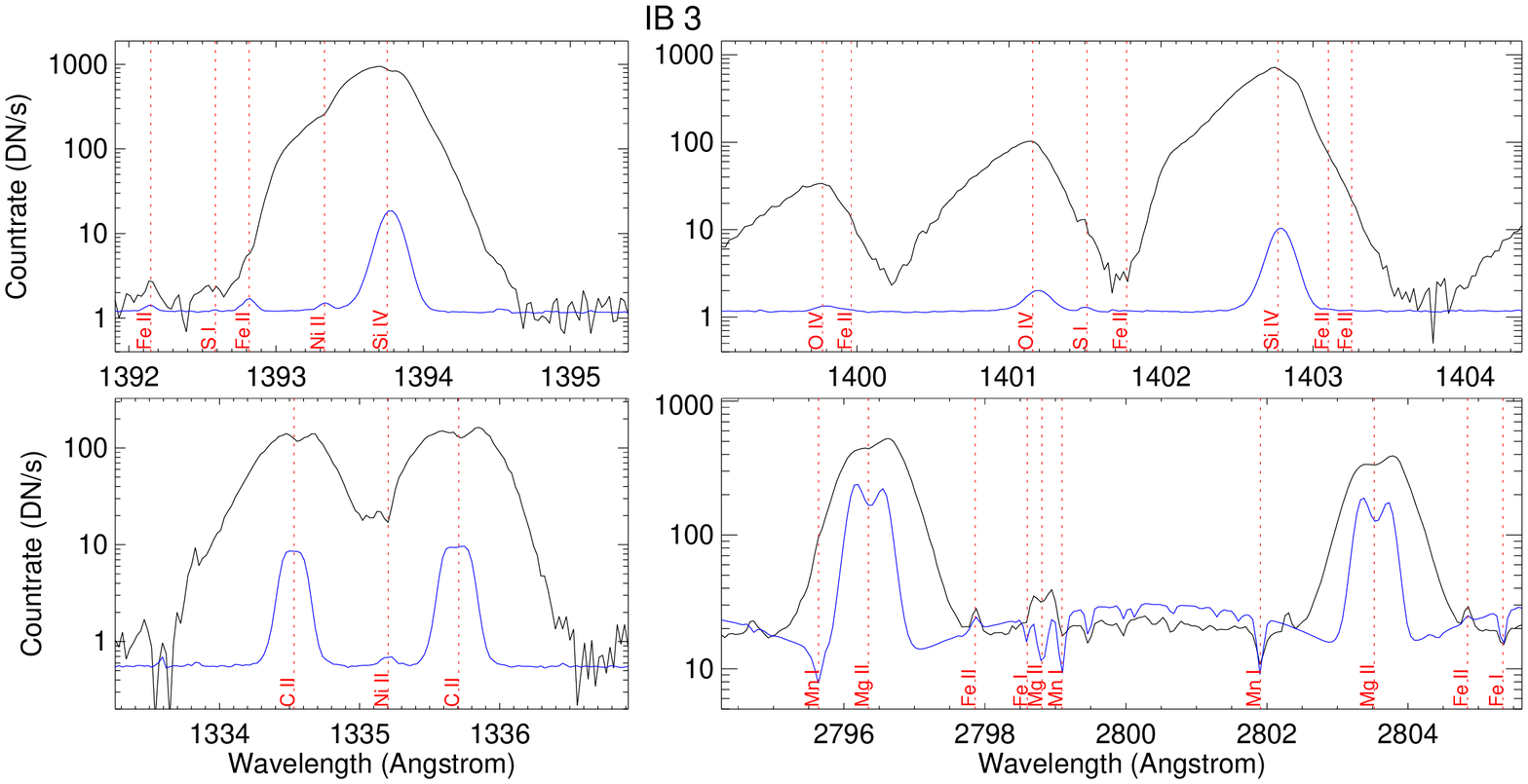}}
\end{minipage}
\begin{minipage}[t]{0.8\textwidth}
{\includegraphics[width=\textwidth]{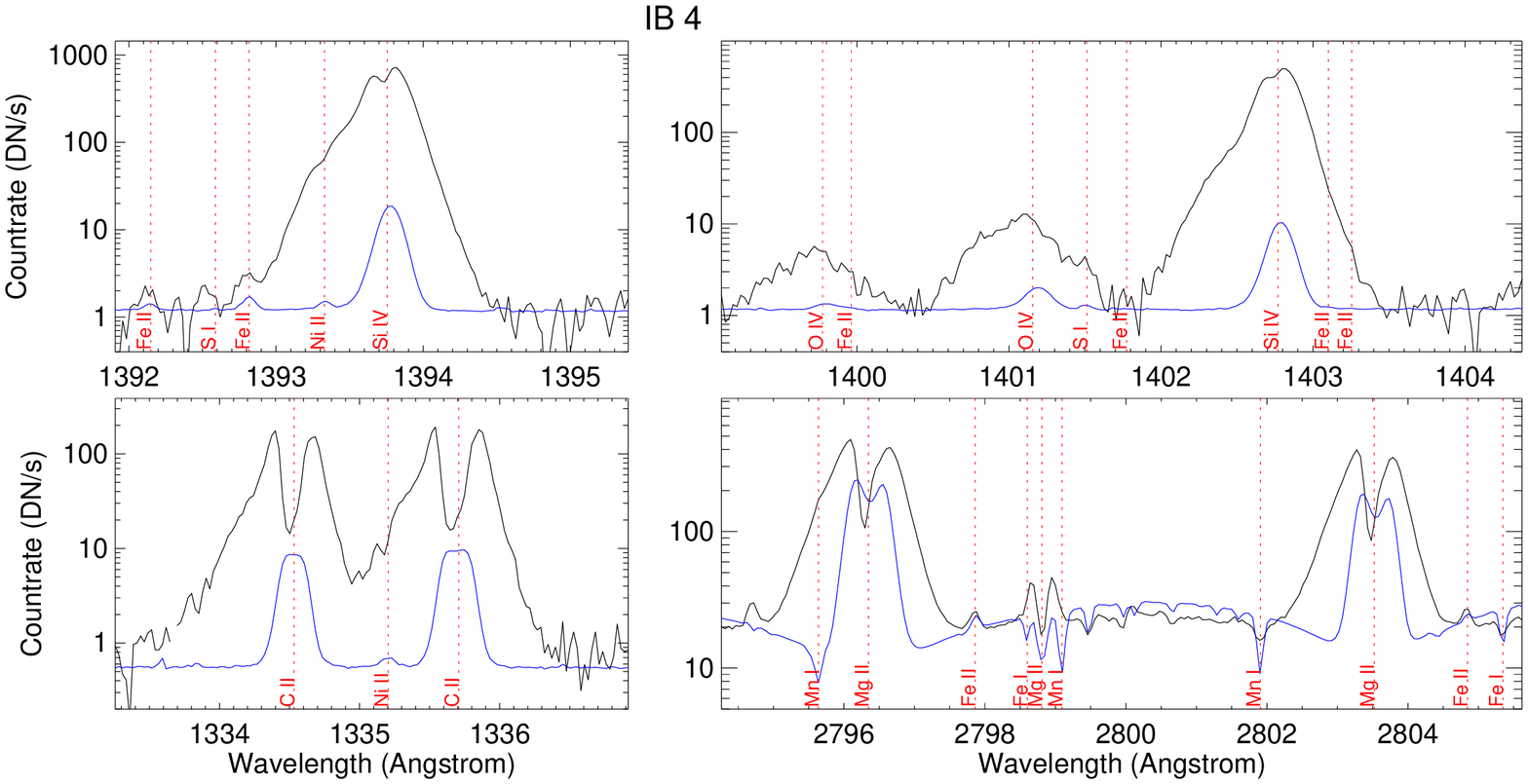}}
\end{minipage}
\begin{minipage}[b]{\textwidth}
\caption{ Same as Figure~\ref{fig.4} but for IBs 3 and 4. } \label{fig.5}
\end{minipage}
\end{figure*}

\begin{figure*}
\centering
\begin{minipage}[t]{0.8\textwidth}
{\includegraphics[width=\textwidth]{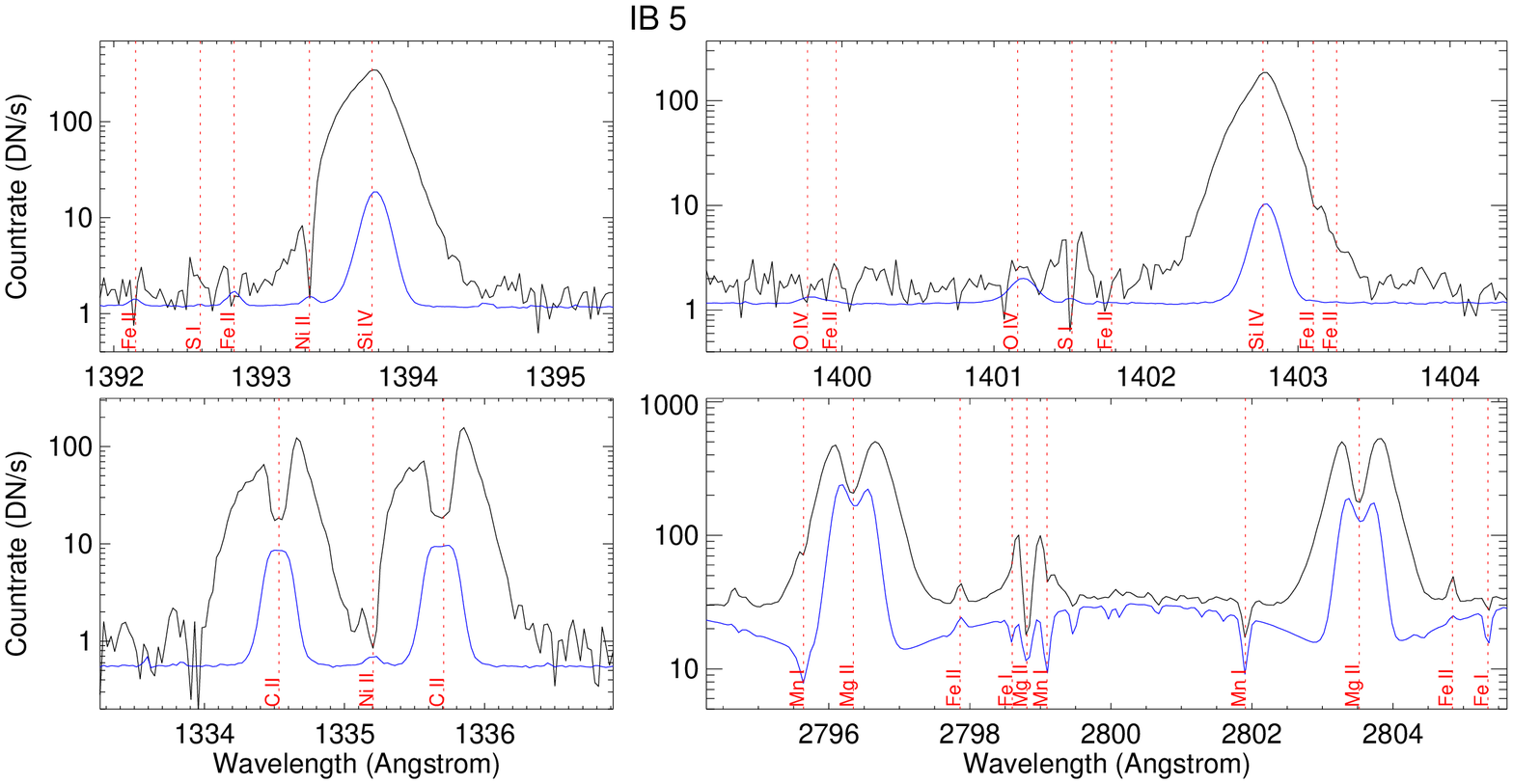}}
\end{minipage}
\begin{minipage}[t]{0.8\textwidth}
{\includegraphics[width=\textwidth]{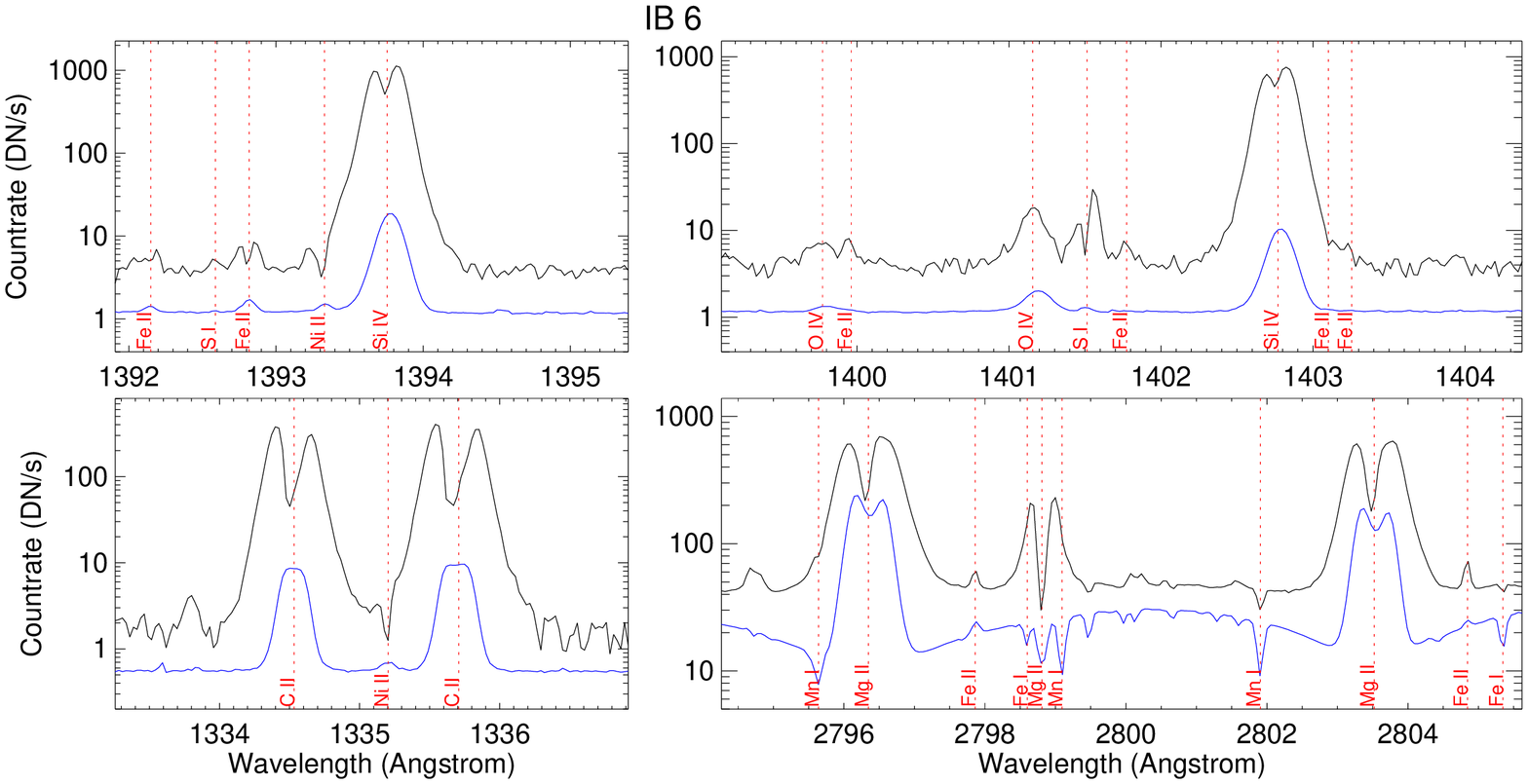}}
\end{minipage}
\begin{minipage}[b]{\textwidth}
\caption{ Same as Figure~\ref{fig.4} but for IBs 5 and 6.  } \label{fig.6}
\end{minipage}
\end{figure*}

\begin{figure*}
\centering
\begin{minipage}[t]{0.8\textwidth}
{\includegraphics[width=\textwidth]{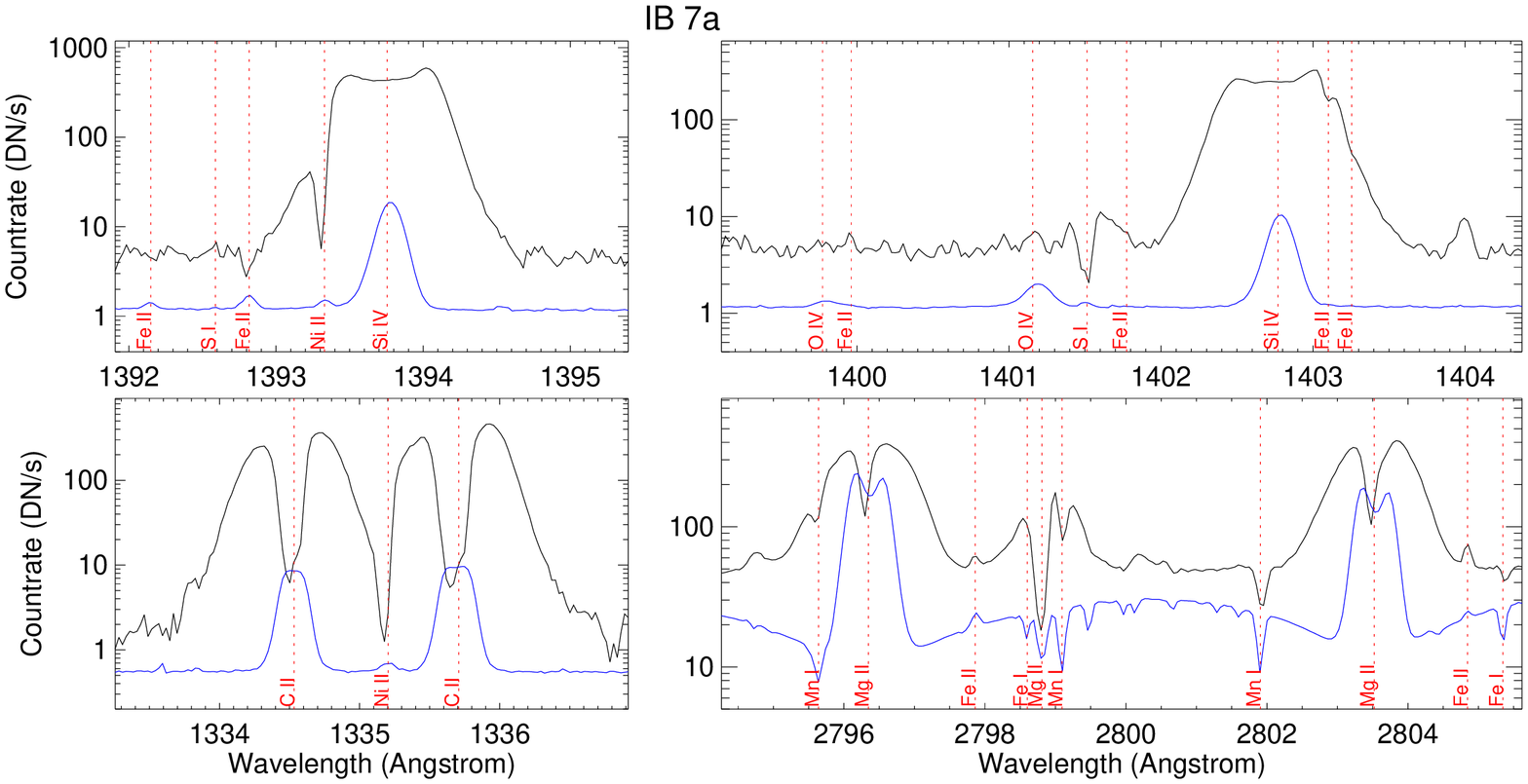}}
\end{minipage}
\begin{minipage}[t]{0.8\textwidth}
{\includegraphics[width=\textwidth]{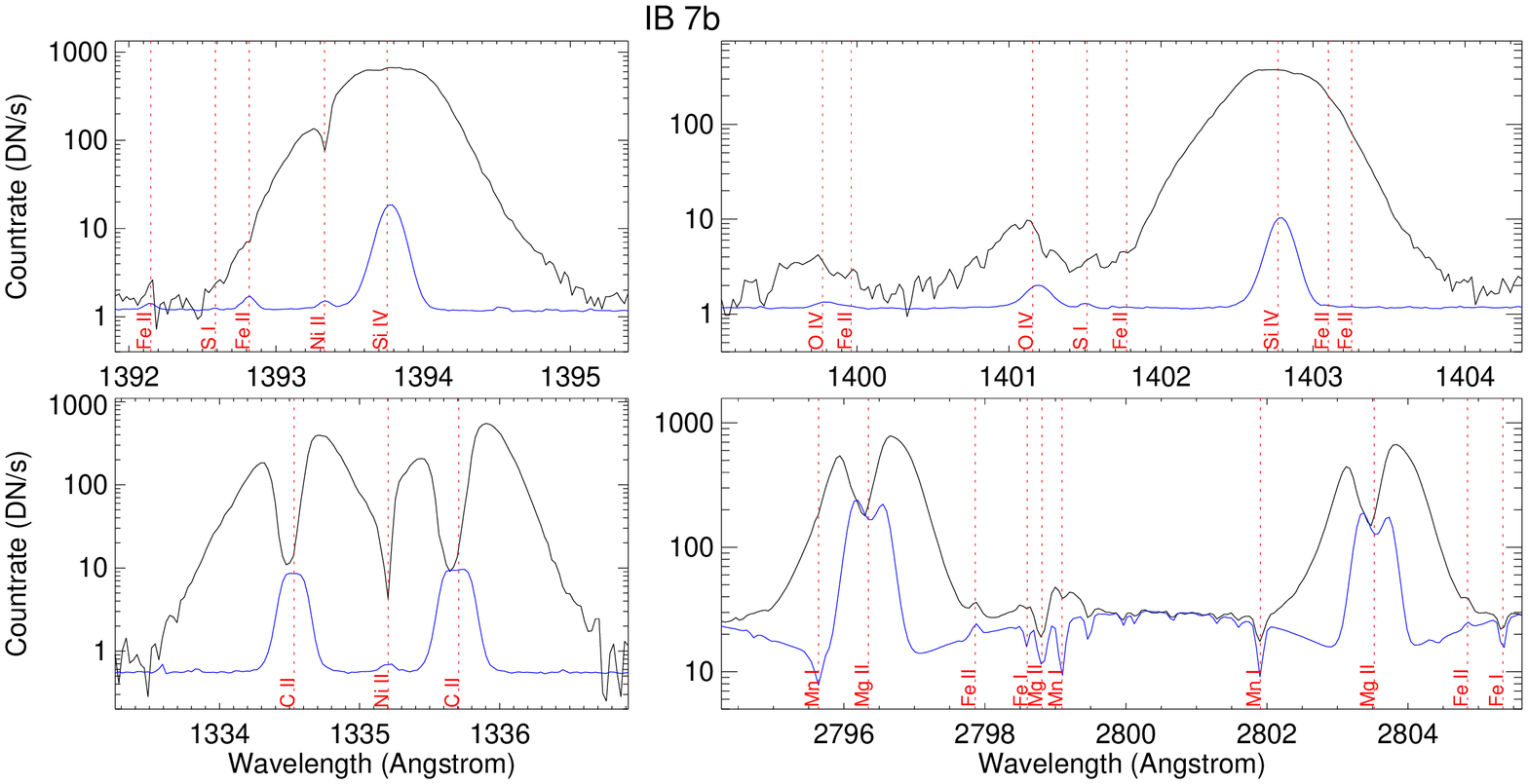}}
\end{minipage}
\begin{minipage}[b]{\textwidth}
\caption{ Same as Figure~\ref{fig.4} but for two different positions in IB 7, as marked in Figure~\ref{fig.3}.  } \label{fig.7}
\end{minipage}
\end{figure*}

\begin{figure*}
\centering
\begin{minipage}[t]{0.8\textwidth}
{\includegraphics[width=\textwidth]{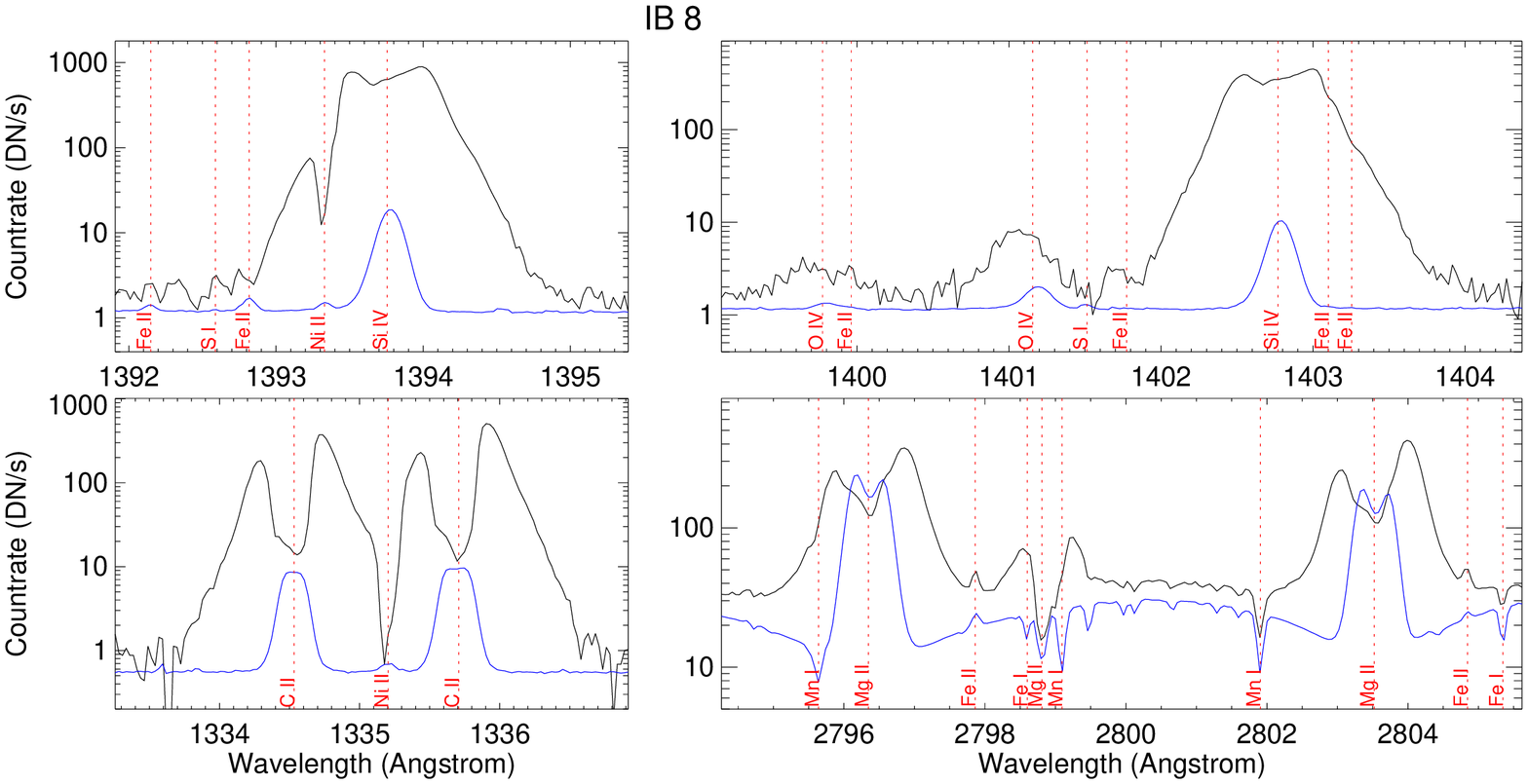}}
\end{minipage}
\begin{minipage}[t]{0.8\textwidth}
{\includegraphics[width=\textwidth]{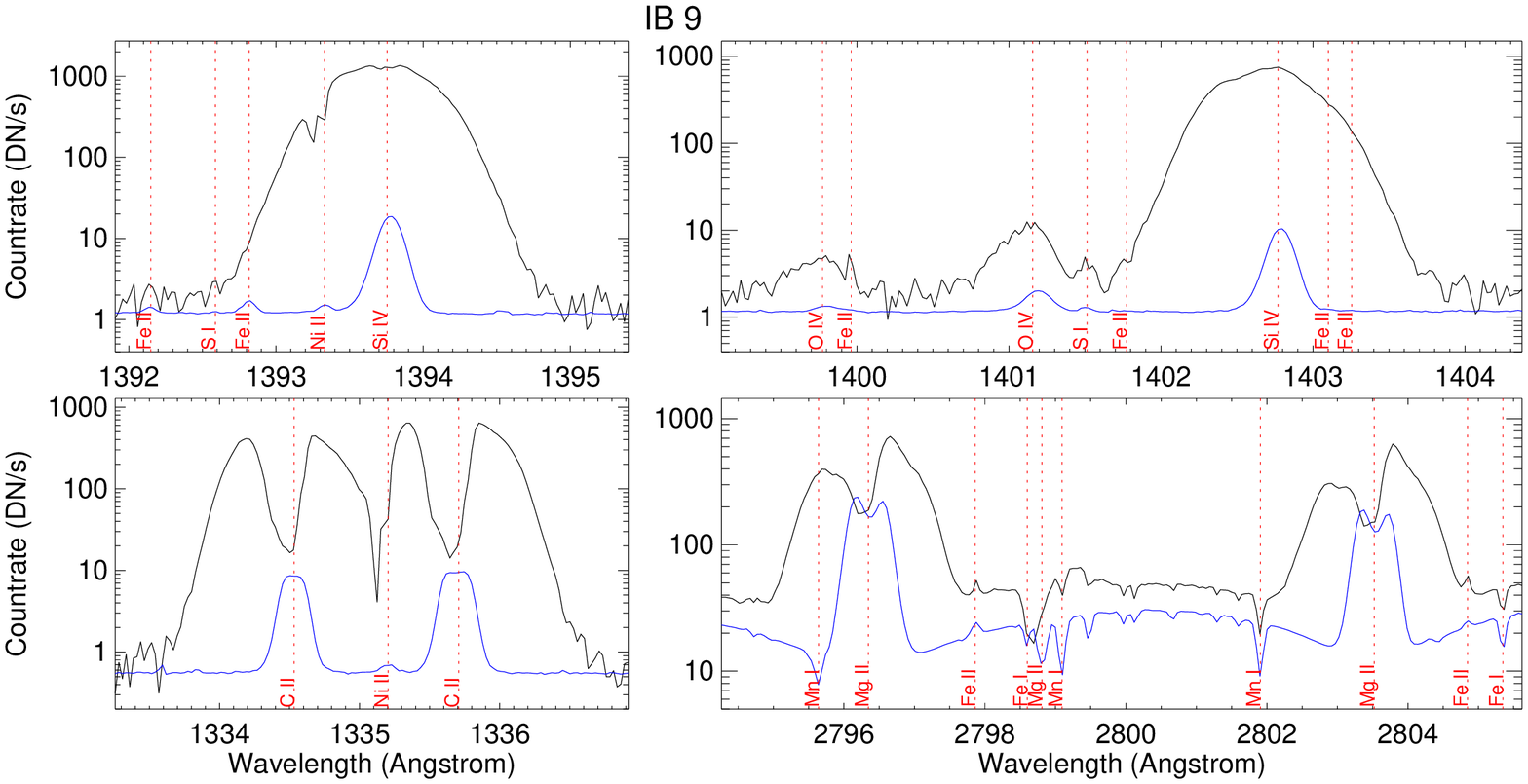}}
\end{minipage}
\begin{minipage}[b]{\textwidth}
\caption{ Same as Figure~\ref{fig.4} but for IBs 8 and 9.  } \label{fig.8}
\end{minipage}
\end{figure*}

\begin{figure*}
\centering
\begin{minipage}[t]{0.8\textwidth}
{\includegraphics[width=\textwidth]{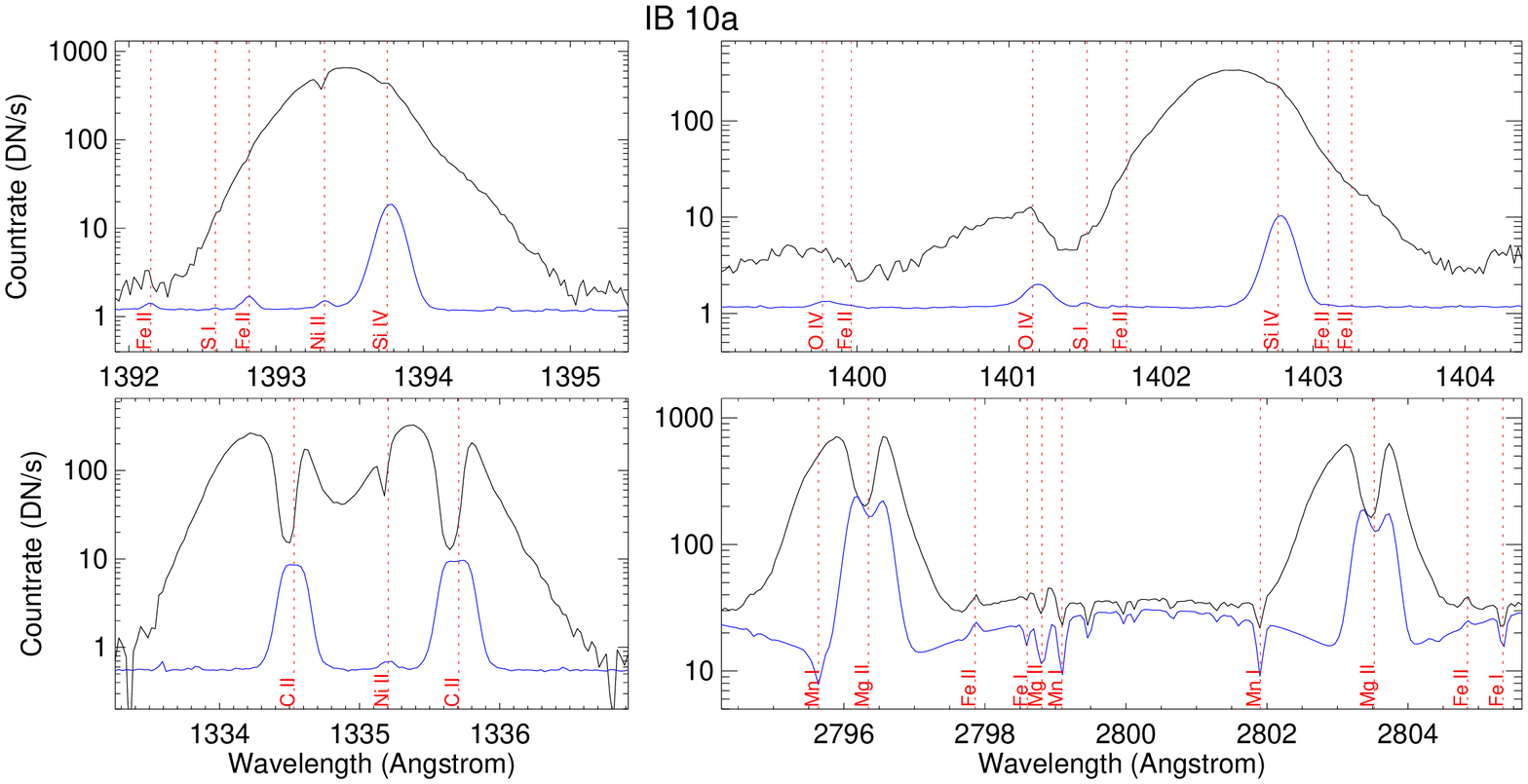}}
\end{minipage}
\begin{minipage}[t]{0.8\textwidth}
{\includegraphics[width=\textwidth]{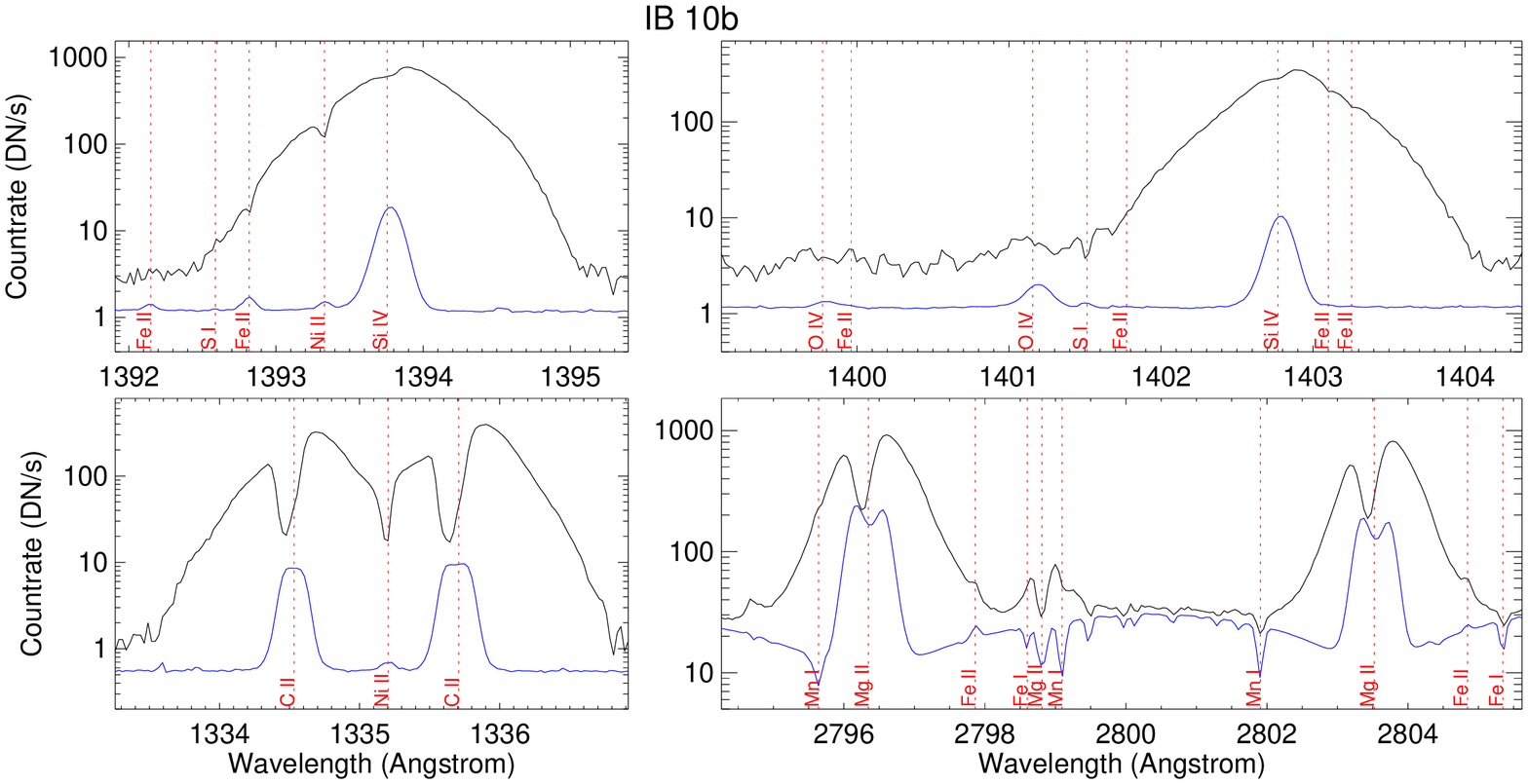}}
\end{minipage}
\begin{minipage}[t]{0.8\textwidth}
{\includegraphics[width=\textwidth]{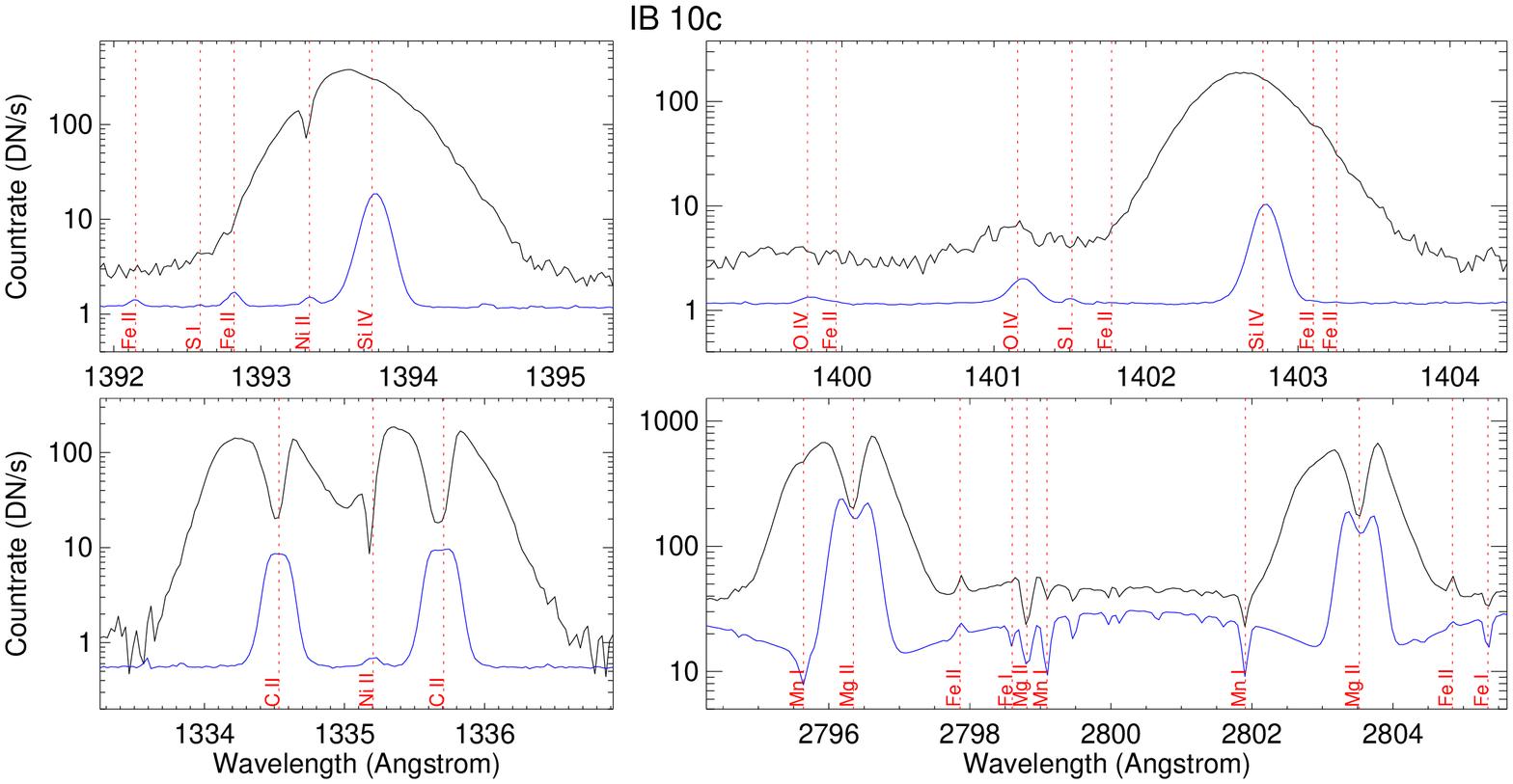}}
\end{minipage}
\begin{minipage}[b]{\textwidth}
\caption{ Same as Figure~\ref{fig.4} but for three different positions in IB 10, as marked in Figure~\ref{fig.3}.} \label{fig.9}
\end{minipage}
\end{figure*}

Typical line profiles of the ten IBs are shown in Figures~\ref{fig.4}--\ref{fig.9}, where the line profiles averaged over a quiet plage region (Solar-X=[--788.5$^{\prime\prime}$ : --775.6$^{\prime\prime}$], Solar-Y=[--222.8$^{\prime\prime}$ : --185.2$^{\prime\prime}$]) are also overplotted for reference. The reference line profiles are also used to perform absolute wavelength calibration. For the Si~{\sc{iv}}~1393.755\AA{}~spectral window, the chromospheric Fe~{\sc{ii}}~1392.817\AA{} and Ni~{\sc{ii}}~1393.330\AA{} lines in the reference spectrum are assumed to have zero Doppler shifts. These cold lines are known to show negligible average velocities in quiet regions. For the Si~{\sc{iv}}~1402.770\AA{}~window, in principle we can assume a zero shift of the S~{\sc{i}}~1401.514\AA{} line. However, this line appears to be weak and also too close to the O~{\sc{iv}}~1401.156\AA{} line. Instead, we calibrate the wavelength for this window by forcing the Si~{\sc{iv}}~1402.770\AA{}~and 1393.755\AA{} lines in the reference spectra to have the same Doppler shift. For the C~{\sc{ii}} window, the Ni~{\sc{ii}}~1335.203\AA{} line is assumed to have zero shift. The wavelength calibration has been confirmed by the similar Doppler shifts of the Ni~{\sc{ii}}~1335.203\AA{} and 1393.330\AA{} absorption lines in the IBs. Absolute wavelength calibration in the Mg~{\sc{ii}} window is much easier since many strong neutral absorption lines are present in the reference spectrum. These neutral lines can be safely assumed to have zero shifts.

As expected, all the IBs show greatly enhanced and broadened profiles of the Si~{\sc{iv}}, C~{\sc{ii}} and Mg~{\sc{ii}} lines, although the Mg~{\sc{ii}} signature appears to be less obvious for some IBs. Obvious enhancement in one or both wings of the Si~{\sc{iv}} lines, which are usually believed to be associated with reconnection outflows  \citep[bidirectional jets or unidirectional jets,][]{Innes1997}, can be clearly identified for most bombs. These line profiles are generally similar to those of transition region explosive events (EEs), which are also believed to result from reconnection \citep[e.g.,][]{Dere1989,Innes1997,Chae1998,Madjarska2004,Ning2004,Teriaca2004,Zhang2010,Huang2014,Gupta2015}. The most distinct difference between EEs and IBs may be the formation height: EEs are formed in the transition region, and IBs are formed lower down as suggested by the chromospheric absorption lines. It is also likely that some previously identified EEs are actually IBs. As demonstrated by \cite{Yan2015}, the narrow absorption features in the far ultraviolet spectra of IRIS may not be unambiguously resolved by previous moderate-resolution instruments such as the Solar Ultraviolet Measurements of Emitted Radiation instrument \citep[SUMER,][]{Wilhelm1995} on board the Solar and Heliospheric Observatory (SOHO). So from spectra taken with these instruments, one can not distinguish between EEs and IBs. We also see obvious absorption features at the cores of the two Si~{\sc{iv}} lines for IBs 4 and 6. The intensity ratio of the two Si~{\sc{iv}} lines is $\sim$1.55 for these two line profiles. This ratio is much smaller than the ratio derived from the reference line profiles, which is 1.93 and close the expected value 2 in optically thin cases. The dip appears to be stronger in the 1393.755\AA{} line. These results suggest that the central absorption feature likely results from self-absorption of the Si~{\sc{iv}} lines rather than bidirectional jets \citep{Yan2015}. It is unclear why the opacity effects become prominent in these IBs. 

Some singly ionized absorption lines are clearly superimposed on the enhanced line profiles of Si~{\sc{iv}} and C~{\sc{ii}}. These lines include Ni~{\sc{ii}}~1335.203\AA{}, Fe~{\sc{ii}}~1392.817\AA{}, Ni~{\sc{ii}}~1393.330\AA{}, Fe~{\sc{ii}}~1403.101\AA{} and Fe~{\sc{ii}}~1403.255\AA{}, with the Ni~{\sc{ii}}~lines being the strongest. They are clearly emission lines in the quiet plage region. Under ionization equilibrium these lines are typically formed at a temperature of log ({\it T}/K)$\approx$4.15, in the upper chromosphere. In IBs they generally reveal a blue shift less than 10~km~s$^{-1}$, although a large blue shift of $\sim$20~km~s$^{-1}$ is found for IB 9. It is believed that these absorption lines result from the largely undisturbed upper chromosphere and they suggest the presence of hotter gas (up to the Si~{\sc{iv}} formation temperature $\sim$8$\times$10$^{4}$ K) below the upper chromosphere \citep{Peter2014,Vissers2015}.

The Mg~{\sc{ii}}~2798.809\AA{} line is a self blend of two lines at 2798.754\AA{} and 2798.822\AA{}.
In most observations they appear as absorption lines, but they come into emission above the limb and in energetic phenomena such as flares \citep[e.g.,][]{Tian2015}. \cite{Pereira2015} undertook a forward
modeling study of this line using three-dimensional radiative magnetohydrodynamic (RMHD) models, and found that it changes from absorption to emission when strong heating occurs in
the lower chromosphere. Interestingly, this line shows enhanced wings in all IBs except IBs 1 and 9. The shape of this line is similar to the Mg~{\sc{ii}}~k line at 2796.347\AA{} and Mg~{\sc{ii}}~h line at 2803.523\AA{} for most IBs. This feature was also noticed by \cite{Vissers2015}.

\subsection{Connection between IRIS bombs and Ellerman bombs}
We now examine the relationship between IBs and EBs. Since EBs are usually defined from H$_{\alpha}$ images, we first examine possible signatures of IBs in the H$_{\alpha}$ core and wing images taken with NVST. It has already been demonstrated that the NVST H$_{\alpha}$ data can be very useful when studying filaments and dynamic events in the lower solar atmosphere \citep[e.g.,][]{Yang2015a,Yang2015b,Bi2015,Yan2015a,Yan2015b,Xue2016}. The H$_{\alpha}$ data acquired during our joint observation is good enough for the identification of EBs. The contrast on the NVST images appear not constant over time due to the varying seeing condition, which has little effect on our identification of EBs since the H$_{\alpha}$ wing brightenings discussed below are mostly very strong. 

From Figure~\ref{fig.3} we find that some IBs appear to be associated with EBs and others are not. IBs 1--4 are clearly not connected to EBs. No obvious brightening can be identified from either the H$_{\alpha}$ wing or core images for IBs 1, 2 and 4. While for IB 3 significant brightening can only be identified from the H$_{\alpha}$ core image. These signatures are not EB signatures. IBs 5, 6 and 7 reveal typical signatures of EBs, which include substantial brightening in both wings of H$_{\alpha}$ and no obvious enhancement in H$_{\alpha}$ line core. The latter suggests that these IBs lie below the chromospheric fibril canopy visible in the H$_{\alpha}$ core images. The wing brightenings clearly exceed those from the much more ubiquitous facular bright points \citep[pseudo-EBs or network bright points,][]{Rutten2013}. We notice that IB 7 reveals as a long bright structure in the Si~{\sc{iv}}~intensity image. Its northern part coincides with significant enhancement in both wings of H$_{\alpha}$. However, only the red wing of H$_{\alpha}$ shows a less prominent enhancement in the southern part. The SJI 1400\AA{} images in the online movies suggest that these two parts might be related to two different brightenings. Despite this possibility, we still regard these two neighboring parts as one IB since they could not be separated in the Si~{\sc{iv}}~intensity image. In the following we select one position at each of the two parts in IB 7 for line profile analysis (positions a and b marked in Figure~\ref{fig.3}). IBs 8, 9 and 10 are possibly EBs. These three IBs all show no obvious core enhancement and clear red wing enhancement. However, the blue wing enhancement is less obvious for IB 8 and not present for IB 9, which may result from the obscuration by spicules. From the online movie associated with Figure~\ref{fig.2}, we see that this is indeed the case for IB 9. The blue wing shows no obvious signature when a spicule is launched nearby. However, the blue wing reveals a significant enhancement after the spicule fades away. No significant blue wing enhancement can be identified for IB 10. We notice that IB 10 is an anemone jet \citep{Shibata2007} revealing an obvious inverted--"Y" morphology (Figure~\ref{fig.2}). The jet reveals as a bright collimated structure in the SJI 1400\AA{} images and a dark one in the H$_{\alpha}$ blue wing images, similar to the quiet-Sun network jets or rapid blueshifted excursions \citep{Tian2014a,Rouppe2015}. We select three positions in IB 10 for detailed analysis of the line profiles: a at the upward jet, b and c at the two legs of the inverted--"Y" structure. The line profiles at positions b and c discussed below suggest that IB 10 is possibly an EB.

 \begin{table*}[]
\caption[]{Characteristics of the ten identified IBs, including H$_{\alpha}$ enhancement, intensity ratio of the Si~{\sc{iv}}~1402.770\AA{} and O~{\sc{iv}}~1401.156\AA{} lines, Mg~{\sc{ii}} k \& h enhancement, enhancement of the NUV continuum between the Mg~{\sc{ii}} k \& h lines, superposition of the Mn~{\sc{i}}~2795.640\AA{} absorption line on the greatly enhanced blue wing of Mg~{\sc{ii}}~k, chromospheric absorption lines such as Ni~{\sc{ii}}~1393.330\AA{} and 1335.203\AA{}, brightening in the AIA 1700\AA{} passband, and broadened S~{\sc{i}}~1401.514\AA{} line with a central reversal. For IBs 7 and 10 characteristics at the positions a and b are shown, respectively. These observational signatures indicate that IBs 1--4 are independent of EBs and that IBs 5--10 are likely connected to EBs. }\label{t1}
\begin{center}
\begin{tabular}{| p{0.2cm} | p{1.4cm} | p{0.7cm} | p{1.7cm} | p{1.6cm} | p{1.6cm} | p{1.4cm} | p{2.5cm} | p{1.5cm} | }
\hline IB & H$_{\alpha}$ enhancement  & Si~{\sc{iv}} /O~{\sc{iv}} & Mg~{\sc{ii}} k \& h enhancement &  NUV continuum enhancement? & Mn~{\sc{i}} absorption on Mg~{\sc{ii}}~k wing? & Ni~{\sc{ii}} absorption & AIA 1700\AA{} brightening & S~{\sc{i}} broadened with reversal?\\
\hline 
 1 & no &  7 & no & no & no & moderate &  diffuse, weak &  no \\
 2 & no &  49 & no & no & no & weak &  diffuse, weak &  no \\
 3 & core &  7 & wings, core & no & no &  weak & diffuse, weak &  no \\
 4 & no &  32 & no & no & no &  weak & diffuse, weak&  no \\
 \hline
5 & wings &  154 & wings & slightly & yes & strong & compact &  yes \\
6 & wings &  69 & wings & yes & yes & moderate &  compact, strong &  yes \\
7 & wings &  498 & wings & yes & yes & strong &  compact, strong &  yes \\
8 & red wing &  129 & wings & yes & yes & strong &  diffuse, weak &  no \\
9 & red wing &  143 & wings & yes & no & strong &  compact, strong &  no \\
10 & red wing &  239 & wings & slightly & yes & moderate &  compact, strong &  yes \\

\hline
\end{tabular}
\end{center}
\end{table*}

\cite{Peter2014} found that the O~{\sc{iv}}~1401.156\AA{} and 1399.774\AA{} lines are absent in their four identified IBs. In our data we find that the O~{\sc{iv}}~lines are absent in some IBs but clearly present in other IBs. We have calculated the intensity ratio of the Si~{\sc{iv}}~1402.770\AA{} and O~{\sc{iv}}~1401.156\AA{} lines for each line profile presented in Figures~\ref{fig.4}--\ref{fig.8}, and found that the ratio is larger than 60 for IBs 5, 6, 7, 8, 9 and 10 (positions b and c). The absence of the O~{\sc{iv}}~line emission has been previously found in sub-arcsecond bright dots above the sunspots' transition region \citep{Tian2014b} and small-scale brightenings at the footpoints of hot loops \citep{Testa2014}. \cite{Olluri2013} found that non-equilibrium ionization can lead to the absence of the O~{\sc{iv}}~lines. The O~{\sc{iv}}~lines can also be greatly suppressed in the presence of non-Maxwellian electron distributions \citep{Dudik2014}. A third explanation for the absence of these forbidden lines is the dominance of collisional de-excitation from the meta-stable level over radiative decay in a high density environment \citep{Feldman1978,Young2015}. Our result appears to support the third scenario because the absent or weak O~{\sc{iv}}~lines are all found in EB-related IBs or possible EB-related IBs. Since EBs are a pure photospheric phenomenon \citep[e.g.,][]{Watanabe2011}, the density should be very high and likely high enough to suppress the radiative decay. On the other hand, IBs 1--4 are not EBs and the strong O~{\sc{iv}}~lines probably suggest a relatively lower density at the formation height. Since the Ni~{\sc{ii}}~1335.203\AA{} and 1393.330\AA{} absorption lines are also present, these IBs should be located below the upper chromosphere and thus are likely in the lower or middle chromosphere. Using the intensity ratio of the O~{\sc{iv}}~1401.156\AA{} and 1399.774\AA{} lines, we have derived the electron densities of these IBs under the assumption of ionization equilibrium \citep[CHIANTI v7.1,][]{Landi2013}. The derived densities for IBs 1--4 are in the range of log ($N_{e}$/cm$^{-3}$) = 11.2--11.9, which also suggests the formation of these IBs in the chromosphere. Thus, our results are not necessarily inconsistent with \cite{Judge2015}, who placed the origin of IBs in the low-middle chromosphere or above.

The Mg~{\sc{ii}}~k and h lines also show distinctly different behavior for the EB-related IBs and other IBs. From Figures~\ref{fig.2} and~\ref{fig.3} we can conclude that these Mg~{\sc{ii}}~lines may also be used to identify EBs. For IBs 5--10, we see significant brightening in the Mg~{\sc{ii}}~k wing (--1.33\AA{} and +1.33\AA{} images are similar and thus summed) but no obvious brightening in the Mg~{\sc{ii}}~k line core. These IBs are exactly the ones connected or possibly connected to EBs. While for IBs 1, 2 and 4, we could not identify any significant brightening from either the wing or core images of Mg~{\sc{ii}}~k. At the edge of IB 3 brightening are seen from both the Mg~{\sc{ii}}~k wing and core images. If we replace Mg~{\sc{ii}}~k by H$_{\alpha}$, these properties would indicate that these events are not EBs, which is exactly what we concluded above. The different Mg~{\sc{ii}}~k and h line profiles for these two types of IBs can also be seen from Figures~\ref{fig.4}--\ref{fig.9}, where we generally see significant enhancement of the Mg~{\sc{ii}}~wings and no dramatic change of the line cores for IBs 5--10. The near ultraviolet (NUV) continuum between the Mg~{\sc{ii}}~k and h lines is also enhanced for IBs 5--10, with the only exception at position b of IB 7. While for IBs 1--4, there is no substantial enhancement of the Mg~{\sc{ii}}~wings beyond $\sim$--1.33\AA{}/+1.33\AA{} from the line cores. In addition, the NUV continuum is even suppressed or only slightly enhanced for these IBs. Since the Mg~{\sc{ii}}~k and h cores sample the chromospheric fibrils and wings are formed lower down, the different behavior mentioned above confirms that IBs 5--10 are likely also EBs formed in the photosphere and that IBs 1--4 are not. The enhanced NUV continuum in IBs 5--10 also support our argument that these IBs are likely generated in the photosphere. Our finding suggests that the Mg~{\sc{ii}}~k and h lines may be used similarly to the H$_{\alpha}$ line for the identification and investigation of EBs. This would open a very promising new window for EB studies since the Mg~{\sc{ii}}~k and h data are routinely acquired in the seeing-free IRIS observations. To elaborate this we plan to perform a more detailed analysis using more coordinated observations between IRIS and NVST in the near future.

We also notice that the chromospheric absorption lines such as Ni~{\sc{ii}}~1393.330\AA{} and 1335.203\AA{}, which are superimposed on the broadened and enhanced wings of the Si~{\sc{iv}} and C~{\sc{ii}} lines, are usually very strong in EB-related IBs. These absorption features are generally much weaker (shallower) for other IBs. This difference may also indicate that the EB-related IBs are formed deeper in the atmosphere, thus experiencing stronger absorption at the wavelengths of these chromospheric line. 

Another interesting feature unique to the EB-related IBs is the superposition of the Mn~{\sc{i}}~2795.640\AA{} absorption line on the greatly enhanced blue wing of the Mg~{\sc{ii}}~k line. This feature appears to be very obvious for IBs 5, 6, 7 (position a), 8 and 10 (positions b and c). Similarly, we also see the Mn~{\sc{i}}~2799.093\AA{} absorption feature superimposed on the enhanced red wing of the Mg~{\sc{ii}}~2798.809\AA{} line for IBs 5, 7 and 10 (position b). All these IBs are connected or possibly connected to EBs, as we mentioned above. Since the Mn~{\sc{i}}~lines are formed in the upper photosphere, their absorption lines superimposed on the greatly enhanced Mg~{\sc{ii}}~lines suggest the formation of these hot IBs below the cooler upper photosphere. This again supports our argument that these IBs are also EBs. \cite{Vissers2015} also found this feature for a few EBs. Similar to us, they also attributed these absorption lines to the foreground upper-photosphere gas above the EBs. These NUV absorption lines generally have no obvious Doppler shift, although a small blue shift of Mn~{\sc{i}}~2795.640\AA{} appears to be present for IBs 7 (position a) and 8. This confirms that the upper-photosphere gas is generally not impacted by these EB-related IBs, which are a pure photospheric phenomenon.

\begin{figure}
\centering {\includegraphics[width=0.47\textwidth]{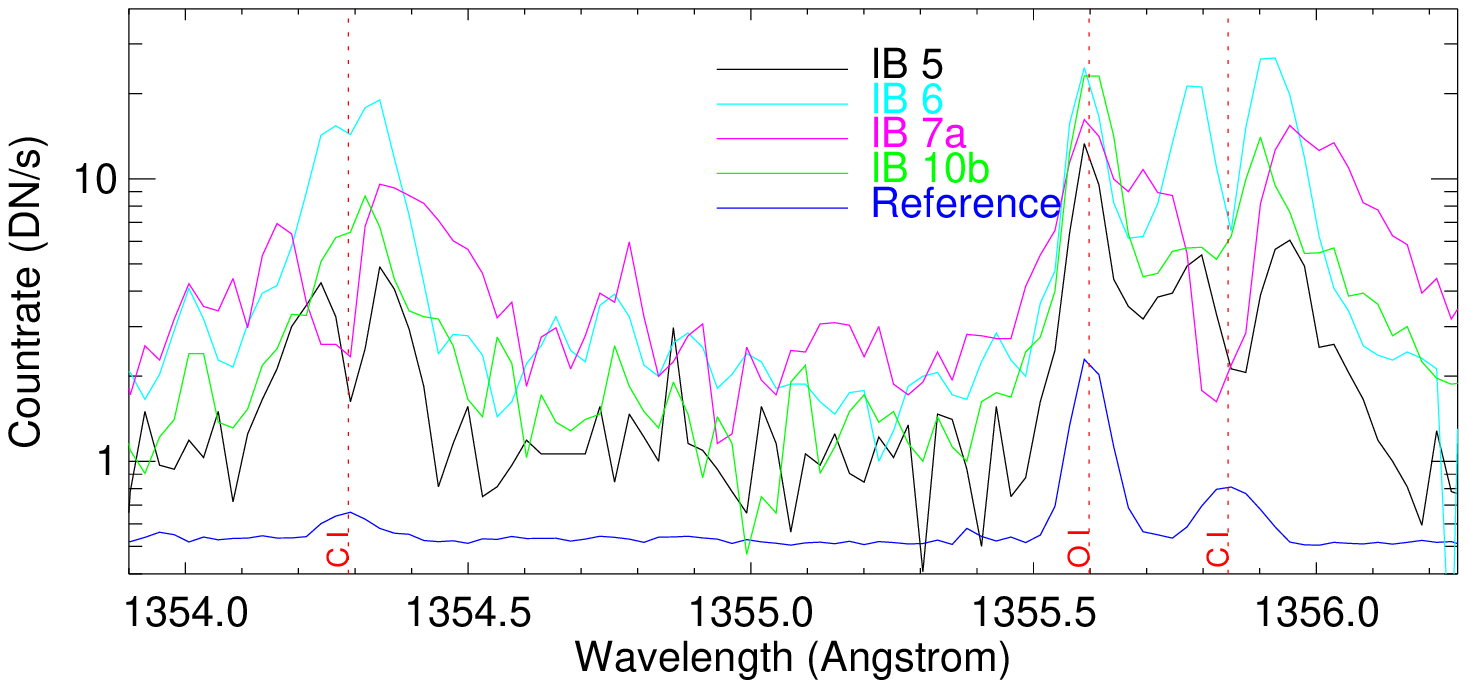}} \caption{ Typical IRIS line profiles of IBs 5, 6, 7 (position a) and 10 (position b) in the
O~{\sc{i}}~1355.598\AA{} window. The blue line profiles represent the reference spectrum obtained in a quiet plage region. Rest wavelengths of
three lines are indicated by the vertical dotted lines. } \label{fig.10}
\end{figure}

From Figures~\ref{fig.6},~\ref{fig.7} and \ref{fig.9} we see that the shape of the S~{\sc{i}}~1401.514\AA{} line profile is similar to those of Mg~{\sc{ii}}~k, Mg~{\sc{ii}}~h and Mg~{\sc{ii}}~2798.809\AA{} for IBs 5, 6, 7 (position a) and 10 (position b), revealing enhanced wings bridged by a dip at the core. In the quiet plage region, this line is simply a weak emission line and its shape is close to Gaussian. We find that the C~{\sc{i}}~1354.288\AA{} and 1355.844\AA{} lines reveal a similar behavior for these IBs. While the optically thin O~{\sc{i}}~1355.598\AA{} line is still very narrow. Figure~\ref{fig.10} shows the profiles of these lines for IBs 5, 6, 7 (position a) and 10 (position b). For wavelength calibration the O~{\sc{i}}~1355.598\AA{} line in the quiet reference spectrum is assumed to have a zero shift. The dramatic change of the S~{\sc{i}}~and C~{\sc{i}}~line profiles in these EB-related IBs is likely caused by the opacity effect. A recent two-cloud model of \cite{Hong2014} reveals an increase of the optical depths when EBs occur, which may result from direct heating in the lower cloud or illumination by enhanced radiation on the upper cloud. The S~{\sc{i}}~and C~{\sc{i}}~line profiles we report here might be related to these processes. 

Finally, from Figure~\ref{fig.3} we notice that the EB-related IBs mostly show strong and compact brightening in AIA 1700\AA{} images. While this is less obvious for other IBs. It seems that IB 8 is not associated with any compact brightening in the AIA 1700\AA{} image shown in Figure~\ref{fig.3}. However, an inspection of the associated online movie suggests that this is because IB 8 was scanned during its decaying phase (at 03:02 UT). In its early phase, at around 02:57 UT, this bomb shows a strong and compact brightening in AIA 1700\AA{}. Since the AIA 1700\AA{} passband samples emission mainly from the upper photosphere, it is not surprising that the chromospheric IBs have little response in the AIA 1700\AA{} passband. We notice that strong EBs have been found to show obvious brightenings in both the 1700\AA{} \citep{Vissers2013} and 1600\AA{} passbands \citep{Qiu2000}, which is consistent with our finding. \cite{Rutten2016} mentioned that the AIA 1700\AA{} and 1600\AA{} channels are also good EB diagnostics because at high temperature they are dominated by the Balmer continuum which shares the H$_{\alpha}$  properties.

We summarize the characteristics of the ten IBs discussed above in Table~\ref{t1}. These observational signatures indicate that IBs 1--4 are independent of EBs and that IBs 5--10 are likely connected to EBs. For IB 7 the characteristics at positions a is shown in the table. As we mentioned above, the northern (around position a) and southern (around position b) parts of IB 7 might be related to two different brightenings in AIA 1700\AA{} images. We find that position a shows all characteristics typical of EB-related IBs. While position b reveals some difference: the NUV continuum is not enhanced and the S~{\sc{i}}~1401.514\AA{} line does not reveal a central reversal. However, the O~{\sc{iv}}~lines are weak and the Mn~{\sc{i}}~2799.093\AA{} absorption feature is superimposed on the enhanced wing of Mg~{\sc{ii}}~2798.809\AA{}. In addition, the Mg~{\sc{ii}}~k \& h wings are enhanced, although not as significant as the enhancement at position a. These characteristics suggest that the southern part is  also possibly connected to an EB. IB 10 appears to be an anemone jet \citep{Shibata2007}. The characteristics at the footpoints (positions b \& c) of the jet, including the absence of the O~{\sc{iv}}~lines, the superposition of the Mn~{\sc{i}}~absorption lines on the enhanced Mg~{\sc{ii}}~wings, the centrally reversed S~{\sc{i}}~line and the intense AIA 1700\AA{} brightening, suggest that it is likely produced in the photosphere and thus is connected to an EB. These features are absent at the tip of the jet (around position a), consistent with the upward flow location higher up in the atmosphere. In Table~\ref{t1} we only show the characteristics at positions b for IB 10.

Our finding that some IBs are also EBs greatly challenges previous modelings of EBs, which almost unexceptionally predict a temperature increase by only a few hundred to $\sim$3000 Kelvin in the photosphere or lower chromosphere \citep[e.g.,][]{Fang2006,Isobe2007,BelloGonzalez2013,Berlicki2014,Hong2014}. Our observations suggest the need of models which can produce a much stronger temperature increase in the photosphere.  A recent numerical simulation by \cite{Ni2015} has predicted heating of the chromospheric plasma to $\sim$8$\times$10$^{4}$ K, which might explain our IBs 1--4. It would be interesting to move the reconnection sites down to the photosphere and investigate the associated heating process. By assuming local thermodynamic equilibrium (LTE) for line extinctions during the hot and dense EB onsets, recently \cite{Rutten2016} put the formation of the Si~{\sc{iv}}~lines in a temperature environment of 1--2$\times$10$^{4}$ K. This temperature, although lower than $\sim$8$\times$10$^{4}$ K under the assumption of ionization equilibrium, is still much hotter than that predicted by all non-LTE modeling of EBs.

So far we have concluded that some IBs are connected to EBs and other are not. One may then ask another question: Does every EB correspond to an IB? From our observation it is clear that not all EBs are connected to IBs. It is impossible to examine the IRIS spectra for all EBs since many EBs were not sampled by the IRIS slit. In addition, sometimes the distinction between EBs and magnetic concentrations (network bright points) is not easy due to the varying seeing condition. Nevertheless, a rough examination of the H$_{\alpha}$ wing images suggests that the IRIS slit crossed $\sim$30 possible EBs, among which only six show typical IB-type line profiles. So it seems that only a small fraction of the EBs are heated to IB temperatures. The other EBs appear not efficiently heated, and thus do not show IB signatures in the Si~{\sc{iv}}~lines. We notice that these events usually also display wing enhancement in the Mg~{\sc{ii}}~k \& h lines, although for many of them the enhancement appears to be weaker compared to that for IB-related EBs.  In the near future we plan to do more coordinated observations between IRIS and NVST, and perform a detailed comparison between the unbinned Mg~{\sc{ii}}~and H$_{\alpha}$ data for EB detection.

\section{Summary}

Using IRIS, NVST, AIA and HMI observations of an emerging AR, we have identified ten IRIS bombs (IBs) and investigated their possible connection to Ellerman bombs (EBs). Seven of these IBs are sitting above the magnetic polarity inversion lines, suggesting that they might result from the interaction between strong magnetic fluxes of opposite polarities. We find that IBs are generally not heated to coronal temperatures.

From the H$_{\alpha}$ images, we find that three IBs are also EBs. Another three IBs are possibly EBs. And the remaining four IBs are obviously not EBs. Considering ionization equilibrium this suggests that EBs can be heated to a temperature of $\sim$8$\times$10$^{4}$ K, one to two orders of magnitude higher than the temperature enhancement predicted from modelings of EBs. According to \cite{Rutten2016}, our result would indicate heating of some EBs to only 1--2$\times$10$^{4}$ K, which is still much hotter than that predicted by all non-LTE modeling of EBs. The EB-related IBs generally reveal the following distinct properties: (1) The O~{\sc{iv}}~1401.156\AA{} and 1399.774\AA{} lines are absent or very weak compared to the Si~{\sc{iv}}~1402.770\AA{} line, likely caused by the high density at the formation height of these IBs. (2) The Mn~{\sc{i}}~2795.640\AA{} and 2799.093\AA{} lines reveal as absorption features superimposed on the greatly enhanced wings of Mg~{\sc{ii}}~k and Mg~{\sc{ii}}~2798.809\AA{} lines, suggesting the shielding of these IBs by the upper photosphere. (3) The Mg~{\sc{ii}}~k and h lines show intense brightening in the wings extending to the nearby NUV continuum and no dramatic enhancement in the cores, suggesting that these IBs are shielded by the overlying chromospheric fibrilar canopy; (4) Absorption features corresponding to the chromospheric Ni~{\sc{ii}}~1393.330\AA{} and 1335.203\AA{} lines are very deep; (5) Intense and compact brightenings can be identified from images of the AIA 1700\AA{} passband which samples the upper photosphere. All together, these features point to the formation of the EB-related IBs in the photosphere. Other IBs may be formed in the chromosphere.

We also find that the shape of the S~{\sc{i}}~1401.514\AA{}, C~{\sc{i}}~1354.288\AA{} and 1355.844\AA{} line profiles reveal enhanced wings bridged by a central reversal, similar to those of Mg~{\sc{ii}}~k, Mg~{\sc{ii}}~h and Mg~{\sc{ii}}~2798.809\AA{} for some EB-related IBs. This behavior likely indicates an increase of the optical depths, as expected when EBs occur \citep{Hong2014}.

Among the ten identified IBs, we find an anemone jet (IB 10) revealing an obvious inverted--"Y" morphology. The characteristics at the footpoints (positions b \& c) of the jet suggest that it is possibly generated in the photosphere, thus demonstrating that some anemone jets reported by \cite{Shibata2007} may result from magnetic reconnection in the partially ionized photosphere.

Finally, a comparison between the IRIS and NVST data suggests that the Mg~{\sc{ii}}~k and h lines could be used to investigate EBs similarly to the H$_{\alpha}$ line, which opens a very promising new window for EB studies since the Mg~{\sc{ii}}~k and h data are routinely acquired in the seeing-free IRIS observations.

\begin{acknowledgements}
The H$_{\alpha}$ data used in this paper were obtained with the New Vacuum Solar Telescope in Fuxian Solar Observatory of Yunnan Astronomical Observatory, CAS. IRIS is a NASA small explorer mission developed and operated by LMSAL with mission operations executed at NASA Ames Research center and major contributions to downlink communications funded by ESA and the Norwegian Space Centre. This work was supported by the Recruitment Program of Global Experts of China, NSFC under grants 41574166, 11473064, 41574168 and 41231069, the Specialized Research Fund for State Key Laboratories, and contract 8100002705 from LMSAL to SAO. H.T. and C.M. thank ISSI Bern for the support to the team "Solar UV bursts -- a new insight to magnetic reconnection". We thank Peter Young, Rob Rutten, Hardi Peter, Brigitte Schmieder, Zhong Liu and the anonymous reviewer for helpful discussion and constructive suggestions. 
\end{acknowledgements}

\end{document}